\documentclass[12pt,twoside,english,a4paper,prd,nofootinbib,preprint,showpacs,preprintnumbers,floatfix]{revtex4}
\usepackage{mathptmx}
\usepackage{helvet}
\usepackage{courier}
\usepackage[T1]{fontenc}
\usepackage[a4paper]{geometry}
\usepackage{color}
\usepackage{graphicx}
\usepackage{amssymb}

\makeatletter

\providecommand{\tabularnewline}{\\}

\@ifundefined{textcolor}{}
{%
 \definecolor{BLACK}{gray}{0}
 \definecolor{WHITE}{gray}{1}
 \definecolor{RED}{rgb}{1,0,0}
 \definecolor{GREEN}{rgb}{0,1,0}
 \definecolor{BLUE}{rgb}{0,0,1}
 \definecolor{CYAN}{cmyk}{1,0,0,0}
 \definecolor{MAGENTA}{cmyk}{0,1,0,0}
 \definecolor{YELLOW}{cmyk}{0,0,1,0}
 }

\makeatother

\usepackage{babel}

\begin{document}
\begin{flushright}
MZ-TH/11-21 \\
TTK-11-26   \\
\par
\end{flushright}

\vspace{4mm}

\begin{center}
\textbf{\LARGE How to pin down the CP quantum numbers}
\par\end{center}{\LARGE \par}

\begin{center}
\textbf{\LARGE of a Higgs boson}
\par\end{center}{\LARGE \par}

\begin{center}
\textbf{\LARGE in its tau decays at the LHC}
\par
\end{center}
{\LARGE \par}

\begin{center}
\vspace{6mm}

\par
\end{center}

\begin{center}
\textbf{\large S. Berge$^{*}$%
\footnote{\texttt{\small berge@uni-mainz.de}}, }
\textbf{\large W. Bernreuther$^{\dagger}$%
\footnote{\texttt{\small breuther@physik.rwth-aachen.de}}, }
\textbf{\large B. Niepelt$^{*}$}
\textbf{\large and H. Spiesberger}$^{*}$\textbf{\large }%
\footnote{\texttt{\small spiesber@uni-mainz.de}} 
\par
\end{center}

\begin{center}
$^{*}$ Institut f\"ur Physik (WA THEP), Johannes Gutenberg-Universit\"at,
55099 Mainz, Germany 
\par
$^{\dagger}$ Institut f\"ur Theoretische Physik, RWTH Aachen University,
52056 Aachen, Germany 
\par
\end{center}

\begin{center}
\vspace{17mm}
\textbf{Abstract}
\par
\end{center}

We investigate how the  $CP$ quantum numbers of  a neutral
Higgs boson or  spin-zero resonance $\Phi$, produced at the CERN 
Large Hadron Collider, can be determined  in its $\tau$-pair 
decay mode $\Phi \to \tau^{-} \tau^{+}$. We use a method 
\cite{Berge:2008dr} based on the distributions of two angles 
and apply it to the major 1-prong $\tau$ decays.
We show for the resulting dilepton, lepton-pion, and two-pion final
states that appropriate selection cuts significantly enhance the
discriminating power of these observables. From our analysis we
conclude that, provided a Higgs boson will be found at the LHC, 
it appears feasible to collect the event numbers needed to 
discriminate between a $CP$-even and $CP$-odd Higgs boson and/or 
between Higgs boson(s) with $CP$-conserving and $CP$-violating 
couplings after several years of high-luminosity runs.  

\vspace{35mm}


\newpage{}


\section{\textbf Introduction}

If the Large Hadron Collider (LHC) at CERN will reach its 
major physics goal of discovering a spin-zero resonance, the next step
will be to clarify the question whether this is the standard model (SM)
Higgs-boson or some nonstandard resonance, as predicted by many of 
the presently  discussed new physics scenarios. (For reviews, 
see \cite{Djouadi:2005gi,Djouadi:2005gj,GomezBock:2007hp,Grojean:2007zz,Morrissey:2009tf}.)
Part of the answer to this 
question will  be given by  measuring  the $CP$ quantum numbers of 
such a particle.   There have been a number of proposals and 
investigations on
how to determine these quantum numbers for Higgs-like resonances
$\Phi$, for several  production and decay processes at hadron
colliders, including
Refs.~\cite{Dell'Aquila:1985ve,Dell'Aquila:1988fe,Bernreuther:1993df,Chang:1993jy,Soni:1993jc,Kramer:1993jn,Grzadkowski:1995rx,Bernreuther:1997af,Plehn:2001nj,Buszello:2002uu,Klamke:2007cu,Berge:2008wi,Berge:2008dr,DeRujula:2010ys}. 
(For an overview, see \cite{Accomando:2006ga}.)
It is the purpose of this article to study a method 
\cite{Berge:2008dr} with which one can pin down whether such a state 
$\Phi$ is $CP$-even, $CP$-odd,  or a $CP$-mixture, namely in its 
decays into $\tau$-lepton pairs with subsequent 1-prong decays. 

Our investigations below are applicable to  neutral spin-zero 
resonances $h_{j}$, for instance to the Higgs-boson(s) of the 
standard model and extensions thereof, with flavor-diagonal 
couplings to quarks and leptons as described by the Yukawa Lagrangian 
\smallskip
\begin{equation}
{\cal L}_{Y} = 
- (\sqrt{2}G_{F})^{1/2}
\sum_{j,f} m_{f} 
\left(a_{jf}\bar{f}f + b_{jf}\bar{f} i \gamma_{5}f \right) \, 
h_{j} \, .
\label{Higg-Lagrangean}
\end{equation}
\smallskip
Here $m_{f}$ is the mass of the fermion $f$ and we normalize the 
coupling constants to the Fermi constant $G_{F}$. A specific 
model is selected by prescribing the reduced scalar and 
pseudoscalar Yukawa coupling constants $a_{jf}$ and $b_{jf}$.
In the standard model (SM) with its sole Higgs-boson, $j = 1$ and 
$a_{jf} = 1$, $b_{jf} = 0$. Many SM extensions predict more 
 than one neutral spin-zero state and  the couplings 
(\ref{Higg-Lagrangean}) can have more general values. Two-Higgs 
doublet models, for instance, the nonsupersymmetric type-II models 
and the minimal supersymmetric SM extension (MSSM, see, e.g., 
\cite{Djouadi:2005gi,Djouadi:2005gj,Accomando:2006ga,GomezBock:2007hp}) 
contain three physical neutral Higgs fields $h_{j}$. If  the Higgs sector
of these models is  $CP$-conserving, or if Higgs-sector $CP$ 
violation (CPV) is negligibly small,  then the fields $h_{j}$ describe 
two scalar states, usually denoted by $h$ and $H$, with $b_{jf} = 0$, 
$a_{jf} \neq 0$, and a pseudoscalar, denoted by $A$, with $a_{jf} = 0$, 
$b_{jf} \neq 0$. In the case of Higgs-sector CPV, the mass eigenstates 
$h_{j}$ are $CP$ mixtures and have nonzero couplings $a_{jf} \neq 0$ 
and $b_{jf} \neq 0$ to scalar and pseudoscalar fermion currents. This 
would lead to $CP$-violating effects in the decays $h_{j} \to f{\bar{f}}$ 
already at  Born level \cite{Bernreuther:1993df}. 

In the following, we use the generic symbol $\Phi$ for any of the 
neutral Higgs-bosons $h_{j}$ of the models mentioned above or, in more 
general terms, for a neutral spin-zero resonance. At the LHC, a $\Phi$ 
resonance can be produced, for instance,  in the gluon and gauge boson fusion processes $g g \to \Phi$ and $q_{i} q_{j} \to \Phi\, q'_{i}q'_{j}$, 
as well as in association with a heavy quark pair,  $t \bar{t} \Phi$ or 
$ b \bar{b} \Phi$. 
Recent studies on Higgs-boson production and decay
  into $\tau$ leptons within the SM and the MSSM include \cite{Baglio:2010ae,Baglio:2011xz}.
Our method for determining
the $CP$ parity of $\Phi$ can be applied to these and to any 
other LHC $\Phi$-production processes. 

The spin of a resonance $\Phi$ can be inferred from the 
polar angle distribution of the $\Phi$-decay products.
In its decays to $\tau$ leptons, $\Phi\to \tau^-\tau^+$,
which is a promising LHC search channel for a number of 
nonstandard Higgs scenarios (see, e.g.\ 
\cite{Djouadi:2005gj,Morrissey:2009tf} and the recent LHC 
searches \cite{Chatrchyan:2011nx,Collaboration:2011rv})
$\tau$-spin correlations induce specific angular distributions 
and correlations between the directions of flight of the 
charged $\tau$-decay products,  in particular an opening 
angle distribution and a $CP$-odd triple correlation and
associated asymmetries \cite{Bernreuther:1993df,Bernreuther:1997af}.
Once a resonance $\Phi$ is discovered, these observables 
can be used to determine whether it is a scalar, a pseudoscalar, 
or a $CP$ mixture. 

The discriminating power of these observables can be exploited fully 
if  the $\tau^{\pm}$ rest frames, i.e., the $\tau$ energies and 
three-momenta can be reconstructed. At the LHC this is possible 
for $\tau$ decays into three charged-pions. With these decay modes 
the $CP$ properties of a Higgs-boson resonance  can be pinned down 
efficiently \cite{Berge:2008wi}. For $\tau$ decays into one charged 
particle the determination of the $\tau^{\pm}$ rest frames is, in 
general, not possible at the LHC. For the 1-prong decays $\tau^\pm 
\to a^\pm$ the  $a^+a^-$ zero-momentum frame can, however, be 
reconstructed. In \cite{Berge:2008dr} two observables, to be 
determined in this frame, were proposed and it was shown, for 
the direct decays $\tau^+\tau^-\to \pi^+ \pi^- {\bar\nu}_{\tau} 
\nu_{\tau}$, that the joint measurement of these two observables 
determines the $CP$ nature of $\Phi$. It will even be possible to  
distinguish (nearly) mass-degenerate scalar and pseudoscalar Higgs-bosons with $CP$-invariant couplings from one or several $CP$ mixtures. 

In  this paper we analyze the $\tau$-pair decay mode of $\Phi$ 
for all major 1-prong $\tau$-decays $\tau^\pm \to a^\pm$ and
investigate how this significantly larger sample can be used 
for pinning down the $CP$ quantum numbers  of $\Phi$ in an 
efficient way. As the respective observables originate from
$\tau$-spin correlations, the  $\tau$-spin  analyzing power
of the charged particle $a$ is crucial for this determination.
The $\tau$-spin  analyzing power of the charged lepton in the 
leptonic $\tau$ decays and of the  charged-pion in the
1-prong hadronic decays $\tau^\pm \to \rho^\pm, a_1^\pm \to \pi^\pm$
is rather poor when integrated over the energy spectrum of the  
respective charged prong. Thus, the crucial question in this 
context is whether experimentally realizable cuts can be
found such that, on the one hand, the $\tau$-spin  analyzing power
of the charged prongs is significantly enhanced and, on the other
hand, the data sample is not severely reduced by these cuts.
We have studied this question in detail and found a positive 
answer.

The paper is organized as follows. In the next section we briefly 
describe the matrix elements on which our Monte Carlo event simulation
is based, and we recapitulate the  two observables with which the
$CP$ nature of a Higgs-boson can be unraveled. In
Sec.~\ref{sec:results} we  analyze in detail the distribution
that discriminates between a scalar and pseudoscalar Higgs-boson,
both for dilepton, lepton-pion, and two-pion final states, for
several cuts. We demonstrate for a set of ``realistic'',
i.e., experimentally realizable cuts that the objectives formulated 
above can actually be met.  This is then also shown for the 
distribution that  discriminates between $\Phi$ bosons with 
$CP$-violating and -conserving couplings. We conclude in 
Sec.~\ref{sect:concl}.

\section{Differential cross section and observables}
\label{sec:LHChpro}

We consider the production of a spin-zero resonance
$\Phi$ -- in the following collectively called a Higgs-boson -- at 
the LHC, and its decay to a pair of $\tau^{\pm}$ leptons:
\begin{equation}
p\, p \to \Phi + X \to \tau^{-} \tau^{+}  + X  \, .
\label{LHC_H_production}
\end{equation}
The decays of $\tau^{\pm}$ are dominated by 1-prong modes with an 
electron, muon, or charged-pion in the final state. We take into 
account the following  modes, which comprise the majority of the
1-prong $\tau$ decays:
\begin{eqnarray}
\tau & \to & l+\nu_{l}+\nu_{\tau} \, ,
\nonumber \\
\tau & \to & a_{1}+\nu_{\tau}\to\pi+2\pi^{0}+\nu_{\tau} \, ,
\nonumber \\
\tau & \to & \rho+\nu_{\tau}\to\pi+\pi^{0}+\nu_{\tau} \, ,
\nonumber \\
\tau & \to & \pi+\nu_{\tau} \, .
\label{eq:tau_decay_channels}
\end{eqnarray}
In the following $a^{\mp} =l^{\mp}, \pi^{\mp}$ refer to the 
charged prongs in the  decays (\ref{eq:tau_decay_channels}).

The hadronic differential cross section $d \sigma$ for the 
combined production and decay processes (\ref{LHC_H_production}), 
(\ref{eq:tau_decay_channels})  can be written as a convolution 
of parton distribution functions and the partonic differential 
cross section  $d\hat{\sigma}$  for $p_1 p_2 \to \Phi \to 
\tau^+\tau^-\to a^+ a'^- + X$ (where $p_1$ and $p_2$ are gluons 
or (anti)quarks):
 \begin{eqnarray}
\label{eq:dsigma_1}
 d\hat{\sigma} & = & 
\frac{\sqrt{2} G_F m_{\tau}^2 \beta_{\tau}}{64\pi^{2}s}
 d\Omega_{\tau}
\overline{\sum}\left|M\left(p_1 p_2 \to \Phi+X \right)\right|^{2}
\left|D^{-1}\left(\Phi\right)\right|^{2}
\mbox{Br}_{_{\tau^{-}\to a'^{-}}}
\mbox{Br}_{_{\tau^{+}\to a^{+}}}
\\[1ex]
 &  & 
 \times \frac{dE_{a'^{-}}d\Omega_{a'^{-}}}{2\pi}
 \frac{dE_{a^{+}}d\Omega_{a^{+}}}{2\pi}
 n\left(E_{a^+}\right) n\left(E_{a'^-}\right)
\nonumber
\\[1ex]
 &  & 
 \times \left(A 
 + b\left(E_{a'^-}\right) 
   {\bf B}^{+} \cdot \hat{\bf q}^{-}
 - b\left(E_{a^+}\right) 
   {\bf B}^{-} \cdot \hat{\bf q}^{+}
 - b\left(E_{a'^-}\right) b\left(E_{a^+}\right) 
   \sum\limits_{i,j=1}^3 C_{ij} \, 
   \hat{q}_{i}^{-} \hat{q}_{j}^{+}\right) \, .
\nonumber 
\end{eqnarray}
Here, $\sqrt{s}$ is the partonic center-of-mass energy,
$\beta_{\tau} = \sqrt{1-4m_{\tau}^2/p_{\Phi}^2}$, 
\begin{equation} 
\label{mat1prop}
\overline{\sum}\left|M\left(p_1 p_2 \to \Phi +X \right)\right|^{2}
\quad \text{and} \quad D^{-1}\left(\Phi \right) 
= 
\left(p_{\Phi}^2 - m_{\Phi}^2 +
\mbox{i}m_{\Phi}\Gamma_{\Phi}^{\rm tot}\right)^{-1} 
\end{equation}
is the spin and color averaged squared production matrix element 
and the Higgs-boson propagator, respectively, with $m_{\Phi}$, 
$p_{\Phi}^{\mu}$ and $\Gamma_{\Phi}^{\rm tot}$ denoting the 
Higgs-boson mass, its 4-momentum and its total 
width\footnote{Equation (\ref{eq:dsigma_1}) holds as long as 
  nonfactorizable radiative corrections that connect the 
  production and decay stage of $\Phi$ are neglected.}. 
The squared matrix element $|T|^2$ of the decay $\Phi \to
\tau^+\tau^- X$, integrated over $X$, is of the form
\begin{equation} \label{matelphitau}
|T|^2 = \sqrt{2} G_F m_{\tau}^2
\left(A +  B^+_i {\hat s^+}_i +  B^-_i {\hat s^-}_i
 + C_{ij}{\hat s^+}_i {\hat s^-}_i\right) \, ,
\end{equation}
where ${\bf\hat s}^{\pm}$ are the normalized $\tau^{\pm}$ spin 
vectors in the respective $\tau^{\pm}$ rest frames. The dynamics 
of the decay  is encoded in the coefficients $A$, $B^{\pm}_i$ and 
$C_{ij}$. Rotational invariance implies that
\begin{equation} \label{bcrotinv}
{\bf B}^{\pm} = 
B^{\pm}\,\hat{\bf k}^{-} \, , \qquad 
C_{ij} 
 = 
c_{1} \delta_{ij} + c_{2} \hat{k}^{-}_{i} \hat{k}^{-}_{j}
+ c_{3}\epsilon_{ijl} \hat{k}^{-}_{l}  \, ,
\end{equation}
where ${\bf k}^-$ $({\bf \hat k}^-)$ is the (normalized) $\tau^-$
momentum in  the $\tau^+\tau^-$ zero-momentum frame (ZMF).
At tree level, $B^{\pm}=0$. (A nonzero absorptive part of the
amplitude, induced for instance by the photonic corrections to 
$\Phi\to \tau \tau$ renders these coefficients nonzero, but the 
effect is very small \cite{Bernreuther:1997af}.) The tree-level 
coefficients $A$ and $c_{1,2,3}$ induced by the general Yukawa 
couplings (\ref{Higg-Lagrangean}) are given in 
Table~\ref{tab:phicouplc} (cf.\ also \cite{Bernreuther:1997af}) 
for a scalar $(b_{\tau}=0)$ and a pseudoscalar $(a_{\tau}=0)$ 
Higgs-boson, $\Phi = H$, $A$, and a $CP$ mixture.
 
\begin{table}[t]
\begin{center}
\begin{tabular}{|c|c|c|c|c|}
\hline 
$\Phi$
    &$A$ 
      &$c_1$
        & $c_2$ 
          & $c_3$
\tabularnewline
\hline 
scalar 
    & $a_{\tau}^2 p_{\Phi}^2 \beta_{\tau}^2/2$ 
      & $a_{\tau}^2 p_{\Phi}^2\beta_{\tau}^2/2$ 
        & $-a_{\tau}^2  p_{\Phi}^2\beta_{\tau}^2$ 
          & $0$ 
\tabularnewline
~pseudoscalar~ 
    & $b_{\tau}^2  p_{\Phi}^2/2$ 
      & $-b_{\tau}^2  p_{\Phi}^2/2$ 
        & $0$ 
          & $0$
\tabularnewline
$CP$ mixture 
    & ~$(a_{\tau}^2 \beta_{\tau}^2+b_{\tau}^2) p_{\Phi}^2/2$~ 
      & ~$(a_{\tau}^2 \beta_{\tau}^2- b_{\tau}^2) p_{\Phi}^2/2$~ 
        & ~$-a_{\tau}^2 p_{\Phi}^2\beta_{\tau}^2$~ 
          & ~$-a_{\tau}b_{\tau} p_{\Phi}^2\beta_{\tau}$~ 
\tabularnewline
\hline
\end{tabular}
\end{center}
\caption{
  Tree-level coefficients of the squared decay matrix element
  (\ref{matelphitau}), (\ref{bcrotinv}) for $\Phi=H$, $A$ 
  (scalar, pseudoscalar) and for a $CP$ mixture. 
\label{tab:phicouplc}
}
\end{table}

We use the narrow-width  approximation for $\tau^\pm$. The branching 
ratios of the 1-prong $\tau$ decays (\ref{eq:tau_decay_channels})
are denoted by ${\rm Br}_{_{\tau^{\pm}\to a^{\pm}}} = \Gamma_{\tau^{\pm}
\to a^{\pm}} / \Gamma_{\tau}^{\rm tot}$. Moreover, the measure 
$d\Omega_{\tau} = d\cos\theta_{\tau}d\varphi_{\tau}$ in (\ref{eq:dsigma_1})
is the differential solid angle of the $\tau^-$ in the Higgs rest frame, 
and $d\Omega_{a^{\pm}} = d\cos\theta_{a^{\pm}}d\varphi_{a^{\pm}}$ and 
$E_{a^{\pm}}$ is the differential solid angle and the energy of the
charged prong $a^{\pm}$ in the $\tau^\pm$ rest frame. Furthermore, the 
functions $n\left(E_{a^{\mp}}\right)$ and  $b\left(E_{a^{\mp}}\right)$ 
encode the decay spectrum of the respective 1-prong polarized $\tau^{\mp}$ 
decay and are defined in the $\tau^{\mp}$ rest frames by 
\begin{eqnarray}
\frac{\mbox{d}\Gamma
  \left(\tau^{\mp}(\hat{\bf s}^{\mp})\to a^{\mp}(q^{\mp})+X\right)}
 {\Gamma\left(\tau^{\mp}\to a^{\mp}+X\right)\, 
  dE_{a^{\mp}}d\Omega_{a^{\mp}}/(4\pi)} 
& = & 
n\left(E_{a^{\mp}}\right) 
\left(1 \pm  
b\left(E_{a^{\mp}}\right)\,\hat{\bf s}^{\mp} \cdot \hat{\bf q}^{\mp} \right) \, ,
\label{eq:dGamma_dEdOmega}
\end{eqnarray}
where $\hat{\bf q}^{\mp}$ is the normalized momentum vector of the 
charged prong $a^{\mp}$ in the respective frame. The function 
$n(E_{a})$ determines the decay rate of $\tau \to a$
while $b(E_{a})$ encodes the $\tau$-spin analyzing power of the
charged prong $a=l,\pi$. We call them spectral functions for short. 
They  are given for the $\tau$-decay modes (\ref{eq:tau_decay_channels})
in Appendix~A. Using the spin-density matrix formalism, the combination 
of (\ref{matelphitau}) and (\ref{eq:dGamma_dEdOmega}) yields, with 
(\ref{mat1prop}), the formula (\ref{eq:dsigma_1}). 

The decay distribution (\ref{matelphitau}) and the coefficients $c_1$,
$c_3$ of Table~\ref{tab:phicouplc} imply that, at the level of the
$\tau^+\tau^-$ intermediates states, the spin observables 
${\bf\hat s}^{+}\cdot{\bf\hat s}^{-}$  and ${\bf\hat k} \cdot 
({\bf\hat s}^{+}\times{\bf\hat s}^{-})$ discriminate between a 
$CP$-even and $CP$-odd Higgs-boson, and between a Higgs-boson with 
$CP$-conserving and  $CP$-violating couplings, respectively 
\cite{Bernreuther:1993df,Bernreuther:1997af}. A the level of the
charged prongs  $a^+a'^-$, these correlations induce a nontrivial  
distribution of the opening angle $\angle({\bf \hat q}^{+}, 
{\bf \hat q}^{-})$ and the $CP$-odd triple correlation ${\bf\hat k}
\cdot({\bf \hat q}^{+}\times {\bf \hat q}^{-})$, as can be read 
off from (\ref{eq:dsigma_1}). The strength of these correlations 
depends on the product $b(E_{a'^-}) b(E_{a^+})$, while $n(E_{a'^-}) 
n(E_{a^+})$ is jointly responsible for the number of  $a^+a'^-$ 
events\footnote{The
  integral $\int n(E_a) b(E_a)dE_a$ determines
  the overall $\tau$-spin analyzing-power of the particle $a$. It
  seems worth recalling that the physics of $\tau$ decays, i.e., 
  the $V-A$ law, has been tested to a level of precision which is much 
  higher than  what is needed for our purposes. Therefore, when 
  comparing predictions with future data, one can use the functions 
  $n(E_a)$ and $b(E_a)$ of the Appendix as determined within
  the standard model.
}.


A direct analysis of experimental data in terms of the kinematic
variables used in the differential cross 
section Eq.\ (\ref{eq:dsigma_1}) is not possible since the momenta 
of the $\tau$ decay products are measured in the laboratory frame 
and the reconstruction of the $\tau^\pm$ and $\Phi$ rest frames is, 
in general, not possible. In Ref.\ \cite{Berge:2008dr} it was 
shown that one can, nevertheless, construct experimentally accessible
 observables with a high sensitivity to the $CP$ quantum numbers of
 $\Phi$. The crucial point is to employ the   zero-momentum frame of 
the $a^+a'^-$ pair. \\
The distribution of  the angle 
\begin{equation}
\varphi^{*} = 
\arccos({\bf \hat{n}}_{\perp}^{*+}\cdot{\bf \hat{n}}_{\perp}^{*-}) 
\label{phistar}
\end{equation}
 discriminates between a $J^{PC}=0^{++}$ and $0^{-+}$ state.
 Here  ${\bf \hat{n}}_{\perp}^{*\pm}$ are normalized impact 
parameter vectors defined in the zero-momentum frame of 
the $a^+a'^-$ pair. 
These vectors can be reconstructed   \cite{Berge:2008dr} from the impact pa\-rameter vectors 
${\bf \hat{n}}_{\mp}$ measured in the laboratory frame by 
boosting the 4-vectors $n_{\mp}^{\mu}=(0,{\bf \hat{n}}_{\mp})$
into the $a'^- a^+$ ZMF and decomposing the spatial part 
of the resulting 4-vectors into their components parallel and 
perpendicular to the respective $\pi^\mp$ or $l^\mp$  momentum. We emphasize 
that $\varphi^{*}$ defined in Eq.\ (\ref{phistar}) is not the 
true angle between the $\tau$ decay planes, but nevertheless, it 
carries enough information to discriminate between a $CP$-even and
$CP$-odd Higgs-boson. \\
The role of  the $CP$-odd and $T$-odd 
triple correlation   mentioned above is taken over by the
             triple correlation ${\cal O}_{CP}^{*} = 
{\bf \hat{p}}_{-}^{*}\cdot({\bf \hat{n}}_{\perp}^{*+}
\times{\bf \hat{n}}_{\perp}^{*-})$ between the impact parameter 
vectors just defined and the normalized $a'^-$ momentum in the 
$a'^- a^+$ ZMF, which is denoted by ${\bf \hat{p}}_{-}^{*}$. Equivalently,
 one can  determine the distribution of 
the angle  \cite{Berge:2008dr} 
\begin{eqnarray}
\psi_{CP}^{*} 
& = & 
\arccos({\bf \hat{p}}_{-}^{*}
\cdot({\bf \hat{n}}_{\perp}^{*+}
\times{\bf \hat{n}}_{\perp}^{*-})) \, .
\label{eq:psi_star_distribution}
\end{eqnarray}

In an ideal experiment, where the energies of the $\tau$ decay 
products $a^\pm$ in the $\tau^\pm$ rest frames  would be  known, 
one could determine the coefficients $A$, $B^{\pm}$, and 
$c_{1,2,3}$  by fitting the differential distribution 
(\ref{eq:dsigma_1}) (using the SM input of the Appendix) to the
data. However, due to  missing energy in the final state, detector 
resolution effects and limited statistics, one has to average 
over energy bins. Moreover, for $a^\mp\neq \pi^\mp$ the function 
$b(E)$ is not positive (negative) definite, see below. Therefore, 
energy averaging can lead to a strong reduction of the sensitivity 
to the coefficients of $b(E)$ in the differential cross section. 
A judicious choice of bins or cuts is therefore crucial to obtain 
maximal information on the $CP$ properties of $\Phi.$ We will
discuss this in detail in the next section.


\section{Results}
\label{sec:results}

The observables (\ref{phistar}) and (\ref{eq:psi_star_distribution})
can be used for the 1-prong $\tau$-pair decay channels of any 
Higgs-boson production process at the LHC. We are interested 
here in the normalized distributions of these variables. If no 
detector cuts are applied, these distributions do not depend on 
the momentum of the Higgs-boson in the laboratory frame; i.e., 
these distributions are independent of the specific Higgs-boson 
production mode.  Applying selection cuts, we have checked for 
some production modes (see below) that, for a given Higgs-boson 
mass $m_{\Phi}\gtrsim 120$ GeV,  the normalized  distributions 
remain essentially process-independent (see also 
\cite{Berge:2008dr}).

For definiteness, we consider in the following the  production of 
one spin-zero resonance $\Phi$ at the LHC ($\sqrt{S} = 14$ 
TeV) in a range of masses $m_{\Phi}$  between 120 and 400 GeV. 
As we employ the general Yukawa couplings (\ref{Higg-Lagrangean}),
our analysis below can be applied to a large class of models, 
including the standard model, 
type-II 2-Higgs doublet models, and  the Higgs sector of the MSSM.
Within a wide parameter range of type-II models, $\Phi$ production is
dominated by gluon-gluon fusion; for large values of the 
parameter $\tan\beta=v_2/v_1$ (where $v_{1,2}$ are the vacuum
expectation values of the two Higgs doublet fields)
the reaction $b{\bar b}\to \Phi$ takes over. (For a recent overview
of various Higgs-boson production processes and the state-of-the-art
of the theoretical predictions, see, e.g., 
\cite{Harlander:2007zz,Dittmaier:2011ti}. 
Higgs-boson production and decay into $\tau^-\tau^+$
 was  analyzed in  the SM and MSSM in
 \cite{Baglio:2010ae,Baglio:2011xz}, 
  taking recent experimental constraints into account.)

For obtaining the results given below we have used the production 
processes  $b{\bar b}\to \Phi$ and $gg \to \Phi$. The reaction chains 
(\ref{LHC_H_production}), (\ref{eq:tau_decay_channels}) were 
computed using leading-order matrix elements only, but our 
conclusions will not change when higher-order QCD corrections 
are taken into account or other Higgs-boson production channels 
with large transverse momentum $p_T^{\Phi}$ are considered. 
Our method can be applied to all production channels,
because no
reconstruction of the Higgs-boson momentum or the $\tau$ momenta
is needed for the determination of the distributions (\ref{phistar}) and 
(\ref{eq:psi_star_distribution}). Therefore, the method is applicable to 
Higgs-boson production with small or large transverse momentum, as
long as the Higgs resonance can be identified in  the $\tau\tau$
events. (For a discussion of the background see the end of Sec.~\ref{suse:combleha}.)

If $p_T^{\Phi}$  is small, the distributions can be
measured as described below. If the $\tau^-, \tau^+$ decay into leptons
or via a $\rho$ or $a_1$ meson, an approximate 
reconstruction of  the Higgs-boson rest frame, as 
 outlined in Sec.~\ref{suse:recoapprox},
will increase the discriminating power of the distributions, because 
appropriate cuts in this frame separate $\tau$-decay
particles with large and small energies.

If $p_T^{\Phi}$  is large,  the reconstruction of 
the  Higgs rest frame can be performed, see \cite{Ellis:1987xu}. 
 With similar cuts as used below, this leads to an even 
 better discriminating power of the 
$\varphi^*$ and $\psi_{CP}^{*} $ distributions.

We have implemented Eq.\ (\ref{eq:dsigma_1}) into a Monte Carlo  
simulation program which  allows us  to study the 
reconstruction of observables in a variety of reference frames  
and to impose selection cuts on momenta and energies.


In Sec.~\ref{suse:leppion} -\ref{suse:combleha} we analyze, for the
various 1-prong final states,  the $\varphi^*$ distributions for
a scalar and pseudoscalar Higgs-boson, i.e., a spin-zero resonance
$\Phi$ with reduced Yukawa couplings $a_{\tau}\neq 0, b_{\tau}=0$
and  $a_{\tau}= 0, b_{\tau}\neq 0$, respectively, to $\tau$ leptons.
For definiteness we choose  $a_{\tau}=1$ and   $b_{\tau}=1$,
respectively. The distribution of the $CP$ angle $\psi_{CP}^{*}$  
is computed in  Sec.~\ref{suse:Hscp} for Higgs-bosons with 
$CP$-violating and $CP$-conserving couplings.


\subsection{Lepton-pion final state: $\tau \tau \to l \pi+ 3\nu$}
\label{suse:leppion}

We start by discussing  the case where the 
$\tau^{-}$ from $\Phi \to \tau^{-}\tau^{+}$ decays leptonically, 
$\tau^{-} \to l^{-} + \bar{\nu}_{l} + \nu_{\tau}$, and the 
$\tau^{+}$ undergoes a direct decay into a pion, $\tau^{+} \to 
\pi^{+} + \bar{\nu}_{\tau}$. The purpose of this section is to 
study the shapes of the $\varphi^*$ distributions for
 scalar and pseudoscalar Higgs-bosons  when  cuts are applied
 to the charged lepton; therefore, no cuts 
are applied at this point to the pion energy and momentum. 

\begin{figure}[t]
\hspace*{-5mm}\includegraphics[scale=0.47]%
{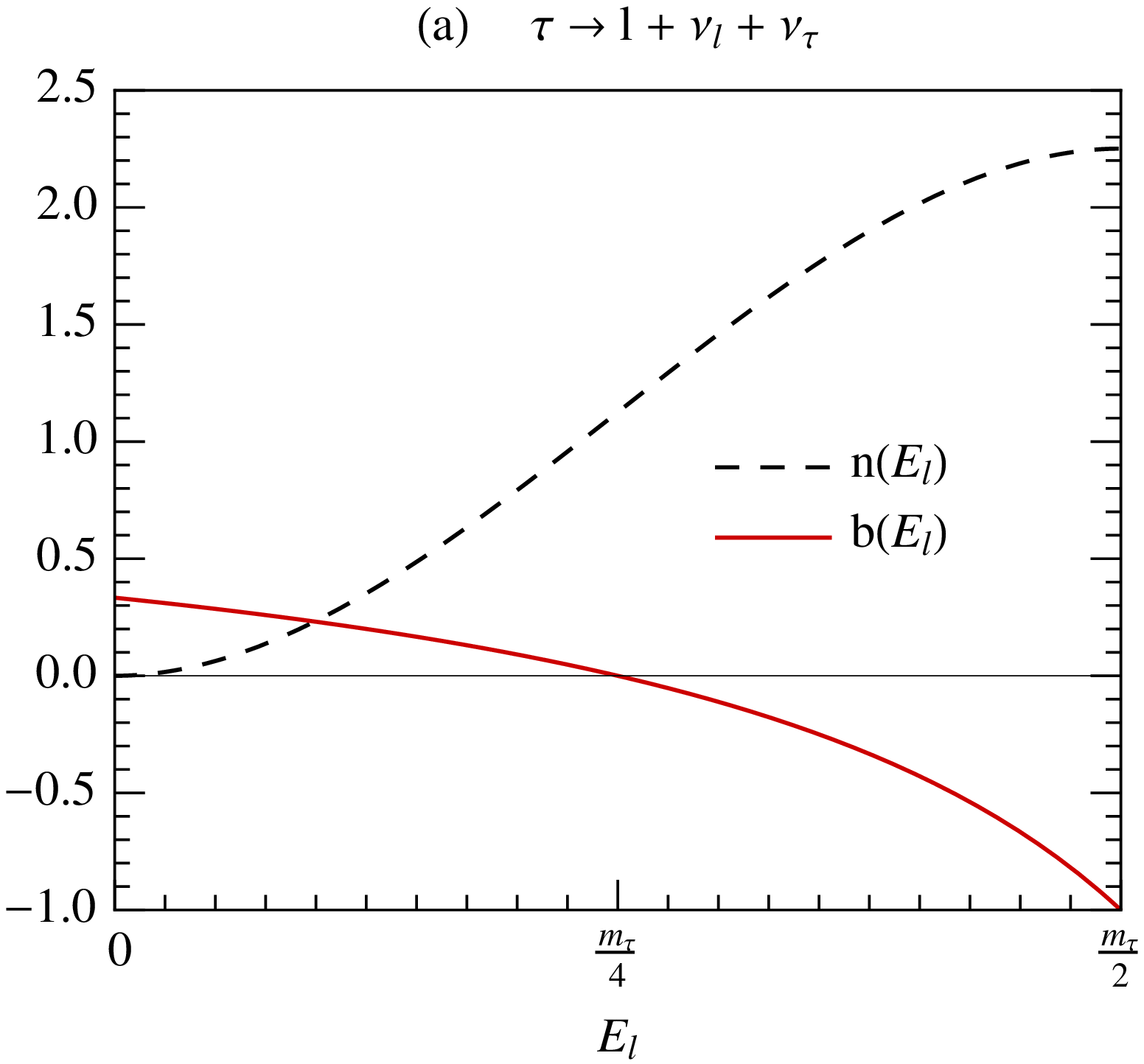}%
\includegraphics[scale=0.47]%
{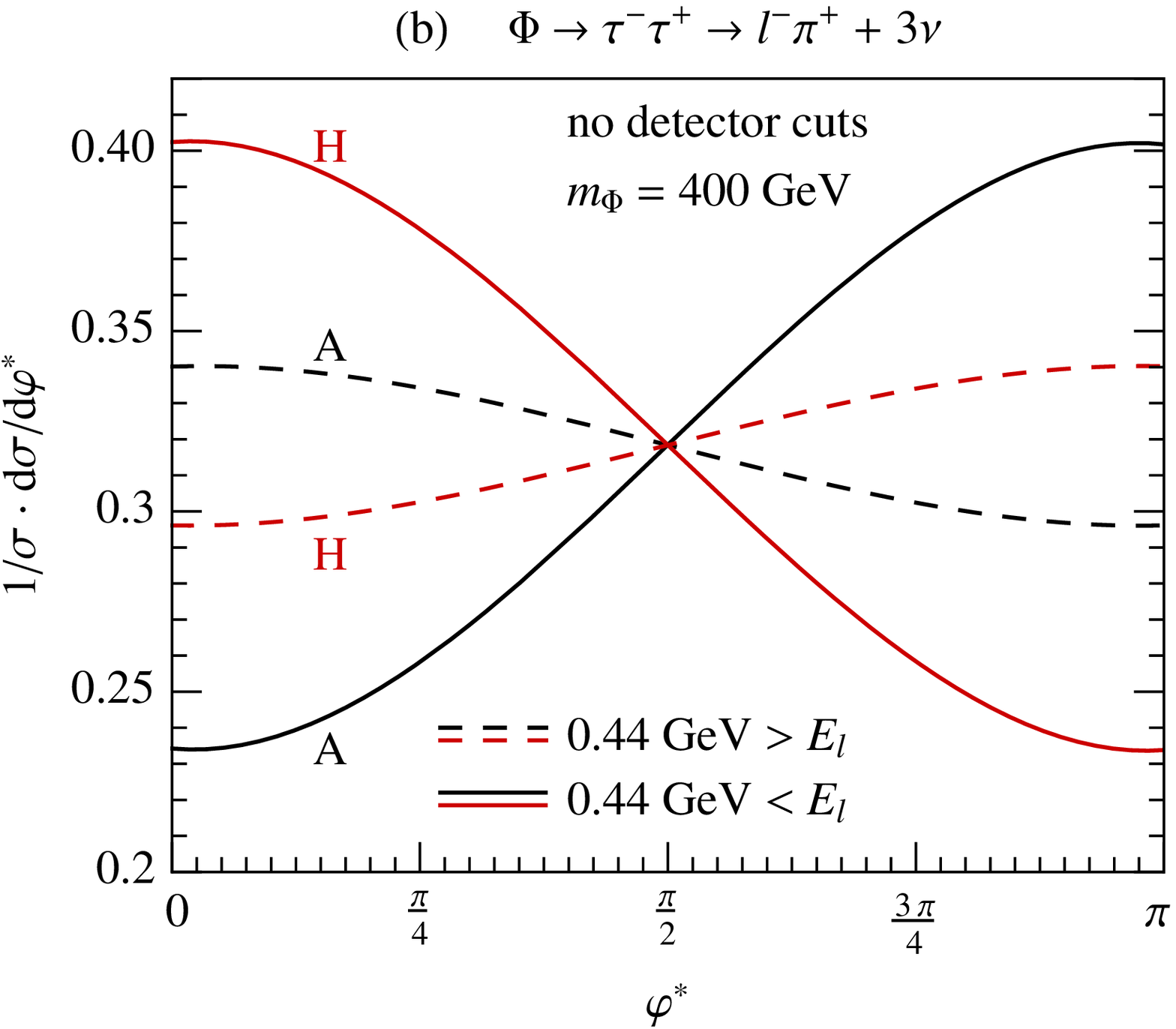}
\caption{(a) The spectral functions  $n(E_{l})$ and 
  $b(E_{l})$, Eq.~(\ref{eq:lep_nx_bx}), for the leptonic 
  $\tau$ decay. The function
  $n(E_l)$ is given in units of ${\rm GeV}^{-1}$. 
  (b) The normalized $\varphi^{*}$ distribution for  $l \pi$ final states
  without selection cuts  in the laboratory frame for a Higgs 
  mass of $m_{\Phi} = 400$~GeV. A cut on the lepton 
  energy  in the $\tau$ rest frame at $m_{\tau} / 4\simeq  0.44$~GeV
  serves to show the  effect of rejecting events where  $b(E_{\,l})$ is
  positive and negative, respectively.
\label{fig:lep_nx_bx}
}
\end{figure}

The charged lepton energy spectrum in the $\tau\to l$ decay is 
determined by the functions $n(E_{l})$ and $b(E_{l})$ given in 
the Appendix, Eq.~(\ref{eq:lep_nx_bx}). These functions are 
shown in Fig.~\ref{fig:lep_nx_bx}(a).
One sees that  the function $b(E_{l})$, which determines the 
$\tau$-spin analyzing power of $l$, changes sign at $E_l = 
m_{\tau}/4$. Therefore also the slope of the resulting 
$\varphi^{*}$ distribution for  $\pi l$ final states changes 
sign at this energy. The optimal way to separate a $CP$-even 
and $CP$-odd Higgs-boson would be to separately integrate over 
the energy ranges $E_l>m_{\tau}/4$ and   $E_l<m_{\tau}/4$. The 
resulting $\varphi^{*}$ distributions for a scalar ($H$, red 
lines\footnote{Color in the electronic version.}) and pseudoscalar 
($A$, black lines) Higgs-boson are shown in 
Fig.~\ref{fig:lep_nx_bx}(b). For the energy range $0 < E_{l} 
< m_{\tau} / 4$, the  $\varphi^{*}$ distribution has  a 
positive  (negative) slope for a scalar (pseudoscalar) Higgs-boson
(dashed curves). For $E_l>m_{\tau}/4$  the slopes change sign and 
the difference between a $CP$-even and a $CP$-odd boson becomes 
more pronounced  (solid curves). However, at a LHC experiment, 
the separation of these two energy ranges is not possible because 
the $\tau$ momenta can not be reconstructed and, therefore,  
the lepton energy $E_l$ in the $\tau$ rest frame can not be 
determined. 

\begin{figure}[t]
\includegraphics[scale=0.47]{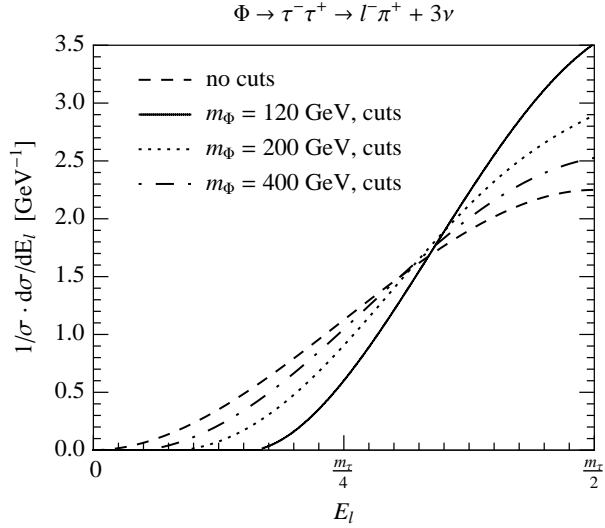}
\caption{
  Normalized lepton energy distribution (in the $\tau$ rest frame) for
  different Higgs-boson  masses, 
  with and without selection cuts
  (\ref{eq:lepton_detector_cuts}). 
\label{fig:lep_El_taurest_detector_cuts}
}
\end{figure}

The difference between the $\varphi^{*}$ distributions for
a scalar and pseudoscalar Higgs-boson  is, however, not 
completely washed out by integrating over the full lepton energy 
range, because both  $n(E_{l})$ and  $b(E_{l})$ have a 
significant energy dependence, see Fig.~\ref{fig:lep_nx_bx}(a). 
The region 
\linebreak[4]$0 < E_{l} < m_{\tau} / 4$ contributes only about 
$19\%$ to the  decay rate $\Gamma_{\tau\to l}$; in addition, 
$b(E_{l})$ is small in this energy range. Therefore, after 
integration over the full $E_{l}$ range, the $\varphi^{*}$ 
distribution is already quite close to the solid lines of 
Fig.~\ref{fig:lep_nx_bx}(b). Moreover, one can suppress
the contribution from the low-energy part of the spectrum by 
imposing a cut on the transverse momentum of the charged lepton 
in the laboratory frame. For the LHC experiments, suitable 
selection cuts on the transverse momentum and the pseudorapidity 
of the lepton in the $pp$ frame are~\cite{Chatrchyan:2011nx,Collaboration:2011rv}: 
\begin{equation}
p_{T}^{l} 
= 
\sqrt{(p_x^l)^2 + (p_y^l)^2}  \ge  20\,\,{\rm GeV} \, ,
\qquad 
|\eta_{l}|  \le  2.5 \, .
\label{eq:lepton_detector_cuts}
\end{equation}
The effect of these cuts on the normalized lepton energy 
distribution in the $\tau^{-}$ rest frame is shown in 
Fig.~\ref{fig:lep_El_taurest_detector_cuts}. Rejecting events 
with small $p^l_{T}$  preferentially removes events with small 
lepton energy in the $\tau$ rest frame.  The effect is more 
pronounced for light Higgs-boson masses where the $\tau$ energy 
is smaller on average. 
For $m_{\Phi} = 120$~GeV, only a small fraction of  $\pi l$ events 
with $E_{l} < m_{\tau}/4$,  about 3.6\%, survives the cuts 
(\ref{eq:lepton_detector_cuts}). For $m_{\Phi}=200$ and 400 GeV 
the corresponding fractions are $9.4\%$ and $14\%$, respectively.  
Events with $E_{l} < m_{\tau} / 4$ that pass the above cuts have 
energies close to $m_{\tau} / 4$, where the function $b(E_{l})$ 
is very small. The resulting $\varphi^{*}$ distribution is almost 
unaffected by contributions with $E_{l} < m_{\tau}/4$, for Higgs 
masses up to $200$~GeV. 
\begin{figure}[t]
\hspace*{-.67cm}
\includegraphics[scale=0.47]%
{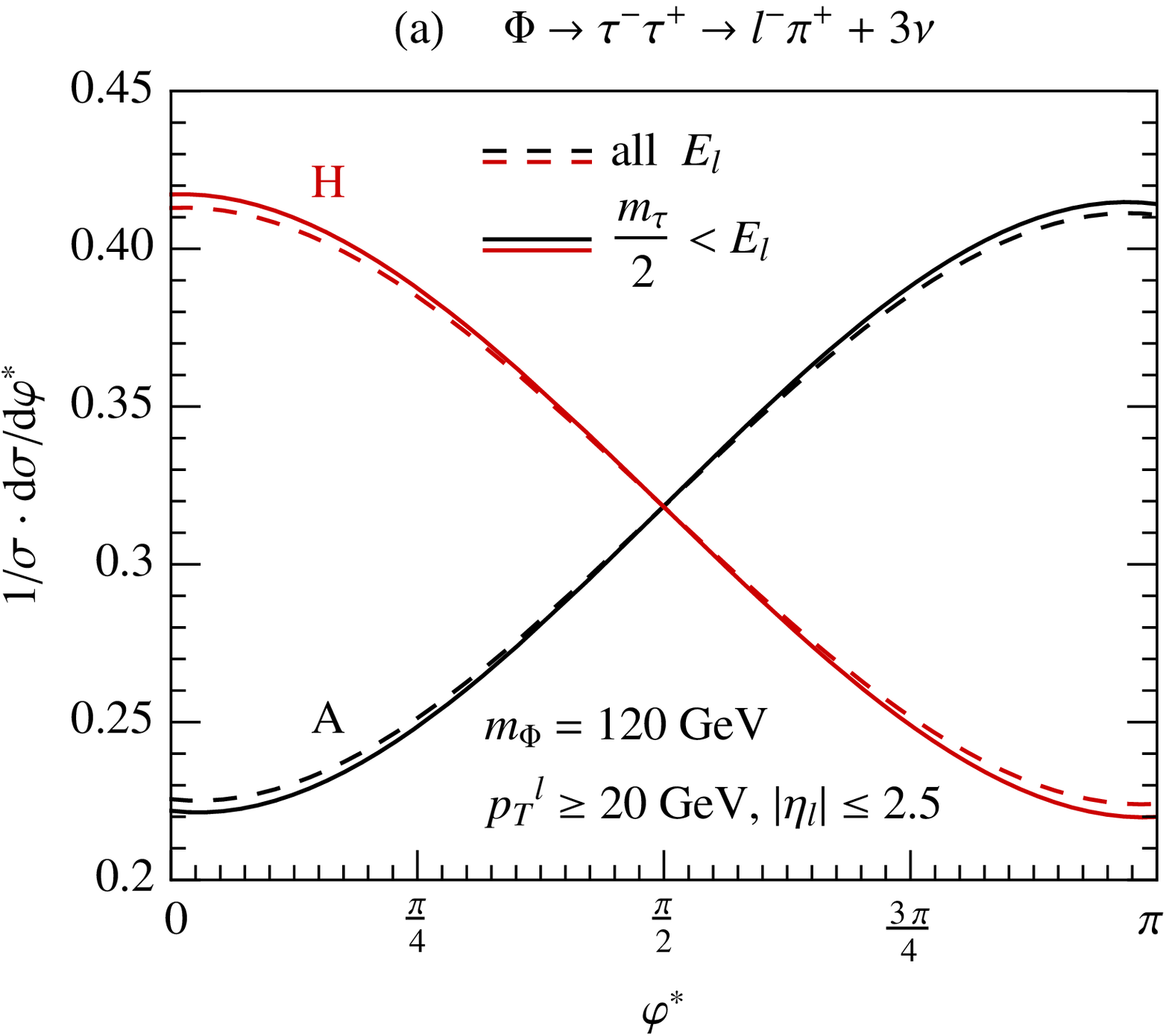}%
\includegraphics[scale=0.47]%
{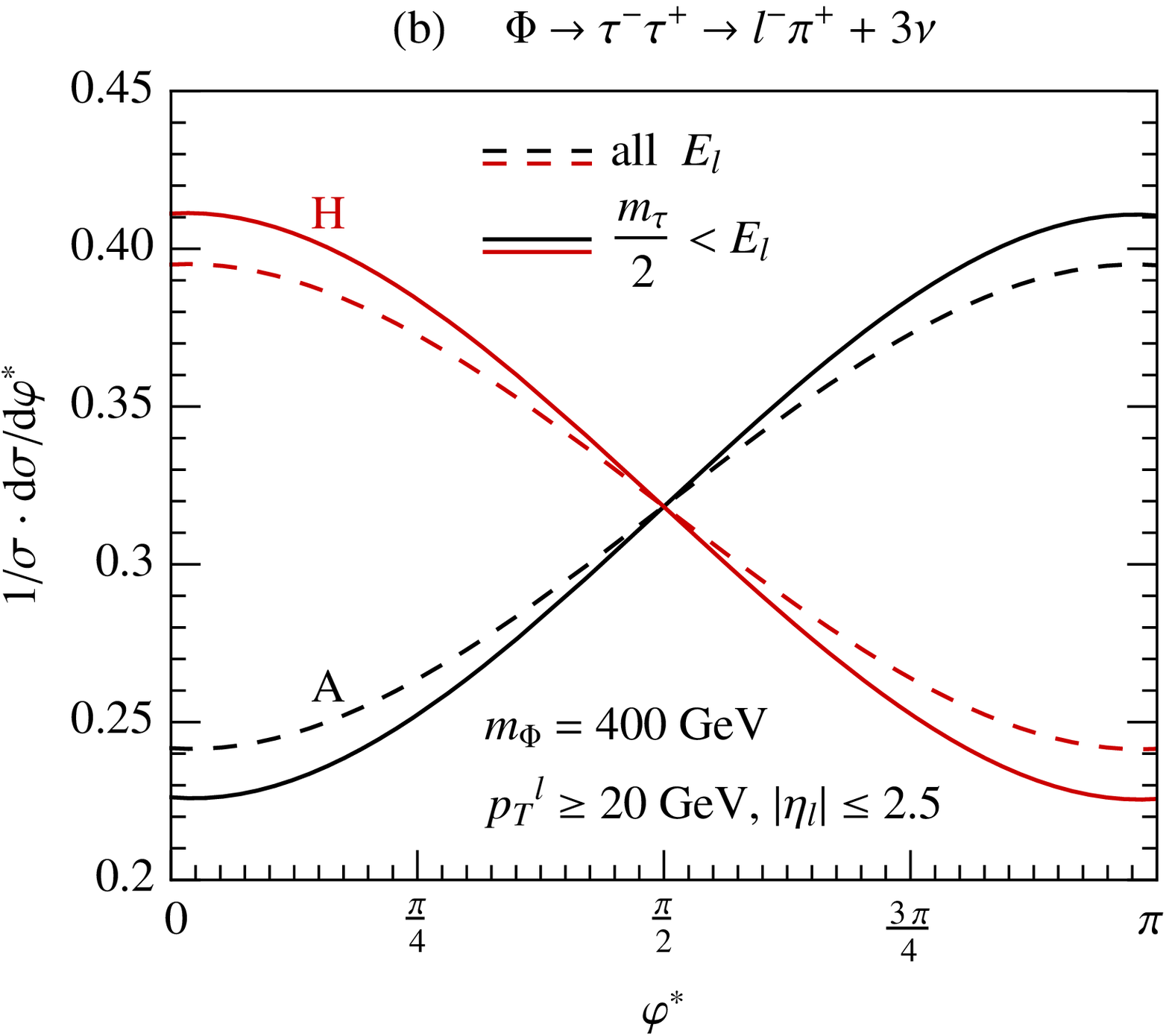}
\caption{
  The normalized $\varphi^{*}$ distributions for $l\pi3\nu$ final 
  states. The solid (dashed) curves show the distribution with the 
  cuts $p_{T}^{l}>20$~GeV and $|\eta_{l}|<2.5$ (without these cuts). 
  (a) $m_{\Phi}=120$~GeV; 
  (b) $m_{\Phi}=400$~GeV.
\label{fig:leppi_phi_pt20eta2.5_Emitheo}
}
\end{figure}
As an example, the $\varphi^{*}$ distributions are  displayed for 
$m_{\Phi}=120$~GeV in Fig.~\ref{fig:leppi_phi_pt20eta2.5_Emitheo}(a) 
and for $m_{\Phi}=400$~GeV in 
Fig.~\ref{fig:leppi_phi_pt20eta2.5_Emitheo}(b). 
From these results we conclude that only for very large 
Higgs-boson masses one can expect to improve the discrimination of 
scalar and pseudoscalar bosons by such a detector cut.

The experimentally relevant case, where in addition also selection 
cuts on the charged-pion are applied, will be discussed in 
Section~\ref{suse:combleha}.


\subsection{Hadronic final states: 
$\tau^{-} \tau^+ \to \{a_{1}^{-}, \rho^{-}, \pi^{-}\} \pi^+$}

Next we analyze the case  where the $\tau^-$ decays to $\pi^-$ either
via a $\rho$ meson, $\tau^{-} \to \rho^{-} + \nu_{\tau} \to \pi^{-} + 
\pi^{0} + \nu_{\tau}$, an $a_1$ meson, $\tau^{-} \to a_{1}^{-} + 
\nu_{\tau} \to \pi^{-} + 2\pi^{0} + \nu_{\tau}$, or directly, $\tau^{-} 
\to \pi^{-} + \nu_{\tau}$, while $\tau^{+}$ undergoes a direct 
2-body decay, $\tau^{+} \to \pi^{+} + \bar{\nu}_{\tau}$. (The 
respective branching ratios are collected in 
Table~\ref{tab:tau_Branching-ratios}). 

The spectral functions  $n(E_{\pi})$ and $b(E_{\pi})$ for the $\rho$ 
and $a_1$ modes are given  in the Appendix and shown in 
Fig.~\ref{fig:a1_rho_nx_bx}. The direct decay mode $\tau^{-} \to 
\pi^{-} + \nu_{\tau}$  is characterized by a constant pion energy 
in the $\tau$ rest frame and has maximal $\tau$-spin analyzing power 
$b=1$. 

\begin{figure}[h]
\hspace*{-.65cm}
\includegraphics[scale=0.47]{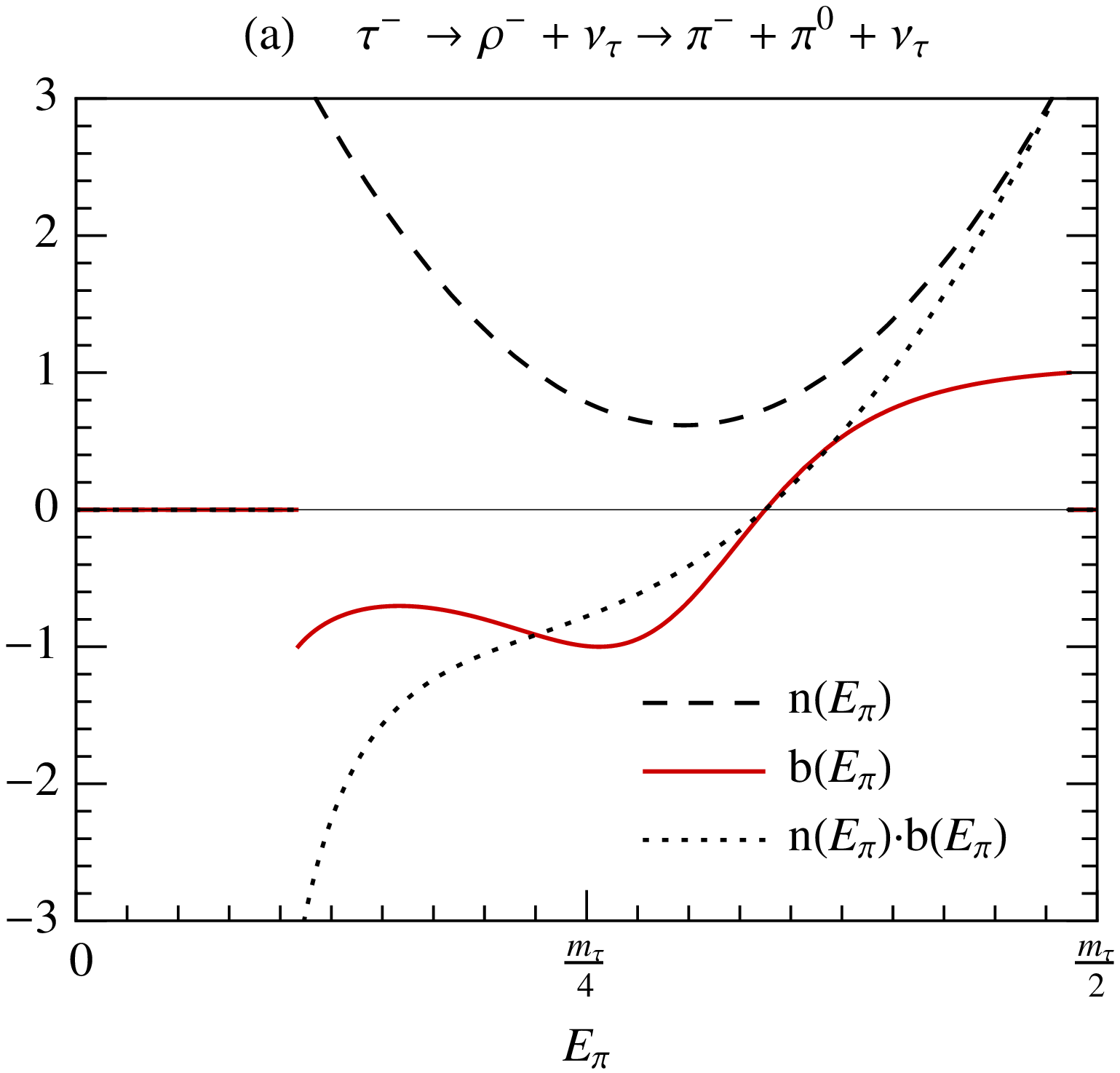}%
\includegraphics[scale=0.47]{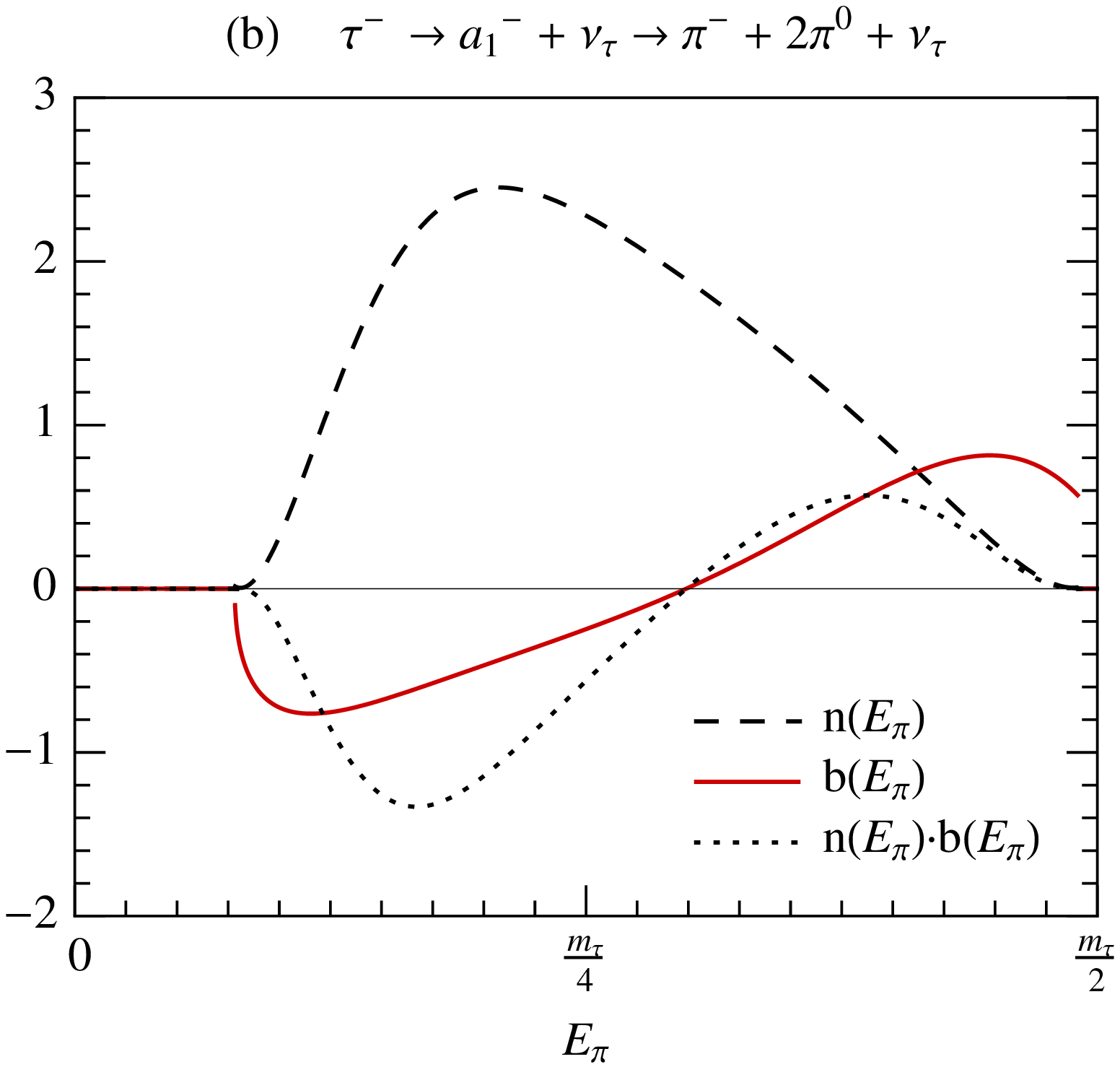}
\caption{
  Charged-pion spectral functions $n(E_{\pi})$ and
  $b(E_{\pi})$ for hadronic $\tau$ decays:
  (a)  $\tau^{-} \to \rho^{-} \nu_{\tau} 
  \to \pi^{-} \pi^{0}  \nu_{\tau}$; 
  (b)  $\tau^{-} \to a_{1}^{-} \nu_{\tau} 
  \to \pi^{-} 2\pi^{0} \nu_{\tau}$. The functions
  $n(E_{\pi})$ and $n(E_{\pi})b(E_{\pi})$   are 
   given in units of ${\rm GeV}^{-1}$.
\label{fig:a1_rho_nx_bx}
}
\end{figure}

\begin{figure}[t]
\hspace*{-.5cm}
\includegraphics[scale=0.47]{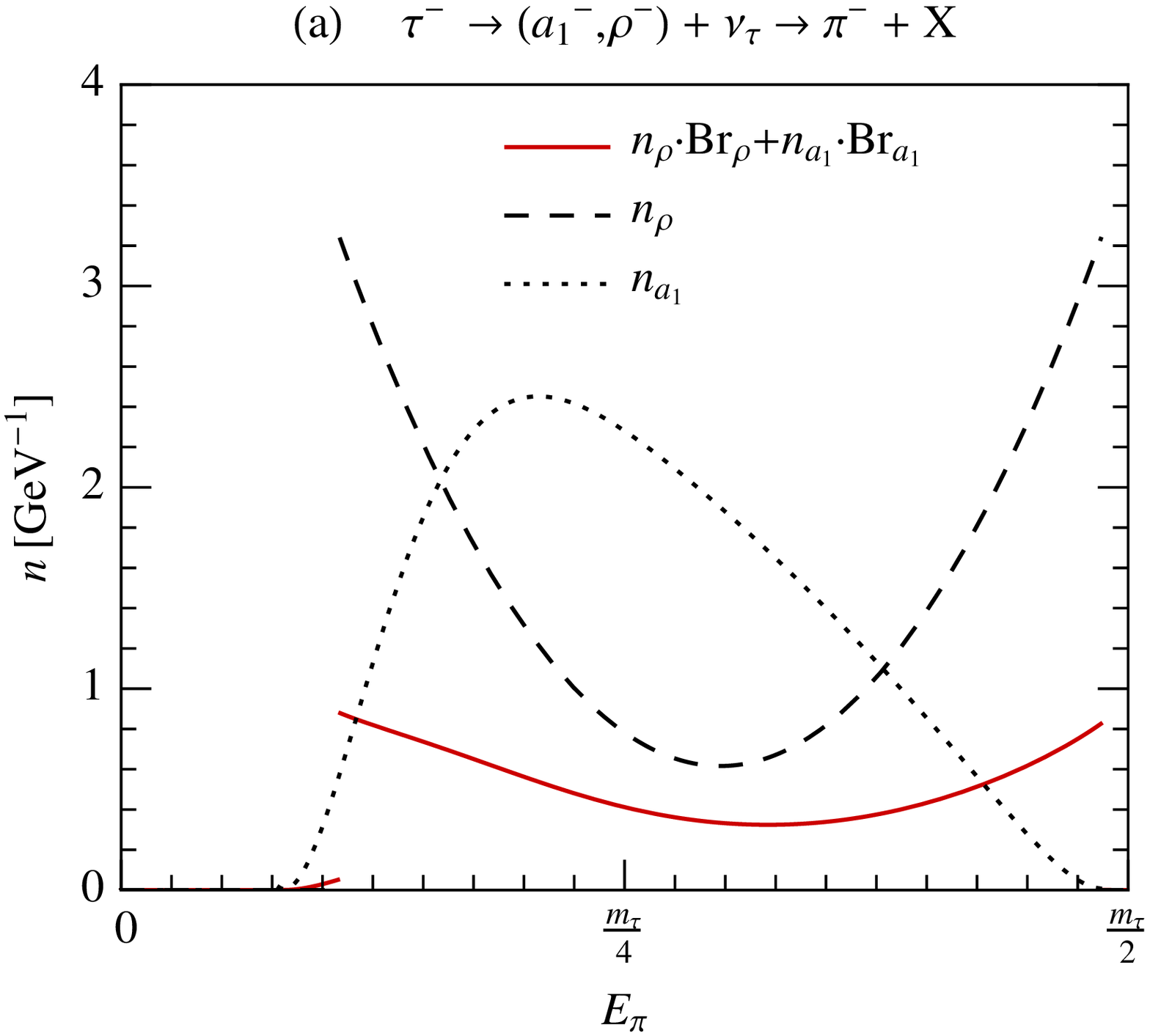}%
\includegraphics[scale=0.47]{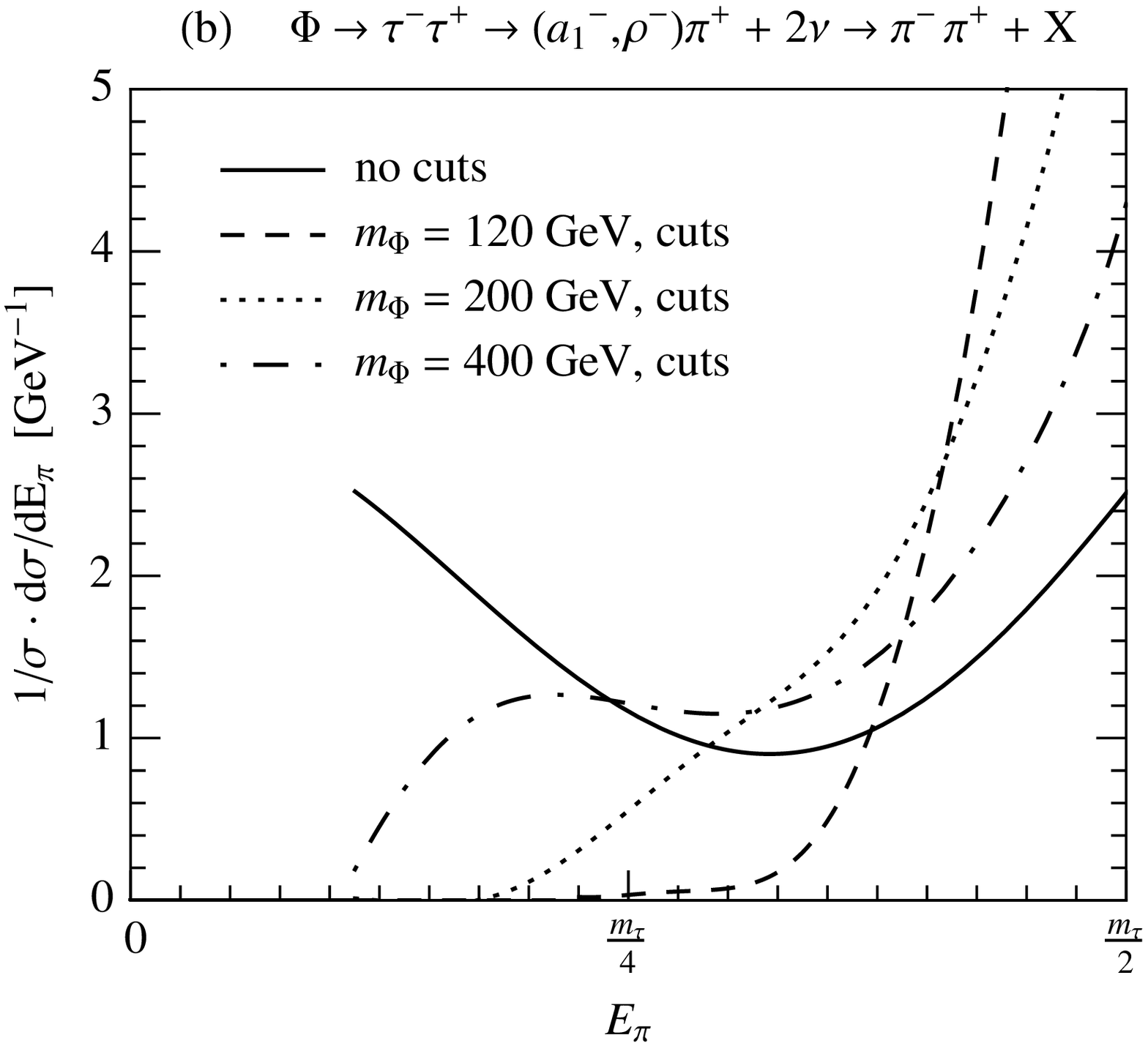}
\caption{
  (a) Spectral function $n(E_{\pi})$ for the combined hadronic decay 
  modes $\tau \to a_{1},\rho \to \pi$ (solid). 
  (b) The distribution of the charged
  pion energy  $E_{\pi}$ for the combined $a_{1}^{-}$ 
  and $\rho^{-}$ decays, for different  Higgs-boson
  masses,  with and without selection cuts. 
\label{fig:a1rho_nx_bx_Epi-taurest}
}
\end{figure}

As in the case of leptonic $\tau$ decay the functions
$b_{\rho}(E_{\pi})$ and $b_{a_1}(E_{\pi})$  change 
sign, at approximately $0.55$~GeV. In contrast to the leptonic 
case, however, contributions from small pion energies are not 
suppressed by small differential rates, as  evidenced by the 
functions $n_{\rho}(E_{\pi})$ and  $n_{a_1}(E_{\pi})$. At the LHC 
one will probably not be able to distinguish between the different
1-prong decay modes into a pion, at least not in an efficient way.
Thus, one has to combine in the Monte Carlo modeling the different 
decay modes, weighted with their branching ratios. The combined 
functions $n(E_{\pi})$ for the decays $\tau \to a_{1}$ and $\tau 
\to \rho$ are shown in Fig.~\ref{fig:a1rho_nx_bx_Epi-taurest}(a). 
The combined distribution is dominated by the $\rho$ decay mode 
because of its larger branching ratio. 
The result (solid red curve) in Fig.~\ref{fig:a1rho_nx_bx_Epi-taurest}(a) 
shows that contributions where $E_{\pi}>0.55$~GeV and $E_{\pi} < 
0.55$~GeV, i.e., where  $b(E_{\pi})$ is positive and negative, 
respectively, are equally important. Without cuts, one would, as 
a consequence, not be able to distinguish between scalar and 
pseudoscalar Higgs-bosons by means of the  $\varphi^{*}$ distribution. 
Therefore we impose the following selection  cuts on the charged-pion 
in the $pp$ laboratory frame, which
 are compatible\footnote{In fact, the
 searches~\cite{Chatrchyan:2011nx,Collaboration:2011rv}
used $p_{T}^{\pi} >  20\,\,{\rm GeV}$;
 but our tighter $p_T$ cut can, of course, always be applied
 in addition to the selected data sample.}  
 with the cuts used by the LHC experiments~\cite{Chatrchyan:2011nx,Collaboration:2011rv}:
\begin{equation}
p_{T}^{\pi}  \ge  40\,\,{\rm GeV} \, ,
\qquad 
|\eta_{\pi}|  \le  2.5 \, .
\label{eq:Had_detector_cuts}
\end{equation}
At this point, the constraints (\ref{eq:Had_detector_cuts}) are
imposed  -- for the purpose of analyzing the effect of these cuts 
-- only on  the $\pi^-$ from $\tau^-$ decays. These cuts will be 
applied to both $\pi^-$ and $\pi^+$ from  $\tau^-$ and $\tau^+$ 
decays, respectively, in Section~\ref{suse:combleha}. 

The impact of these cuts on the  distribution 
$\sigma^{-1}d\sigma/dE_{\pi}$ (where $E_{\pi}$ is the energy of 
the $\pi^-$ in the $\tau^-$ rest frame) for the combined $\tau^{-} 
\to a_{1}^{-}, \rho^{-} \to \pi^{-}$ decay modes is displayed in 
Fig.~\ref{fig:a1rho_nx_bx_Epi-taurest}(b) for several Higgs-boson 
masses between $120$ and $400$~GeV. The solid curve shows the 
distribution without cuts; it does not depend on $m_{\Phi}$. The 
cuts (\ref{eq:Had_detector_cuts}) preferably reject events 
with small $E_{\pi}$.  The effect of these cuts is strong 
for light Higgs-boson masses and still pronounced  for 
$m_{\Phi} \sim 200$~GeV.
 
\begin{figure}[t]
\hspace*{-.1cm}
\includegraphics[scale=0.46]%
{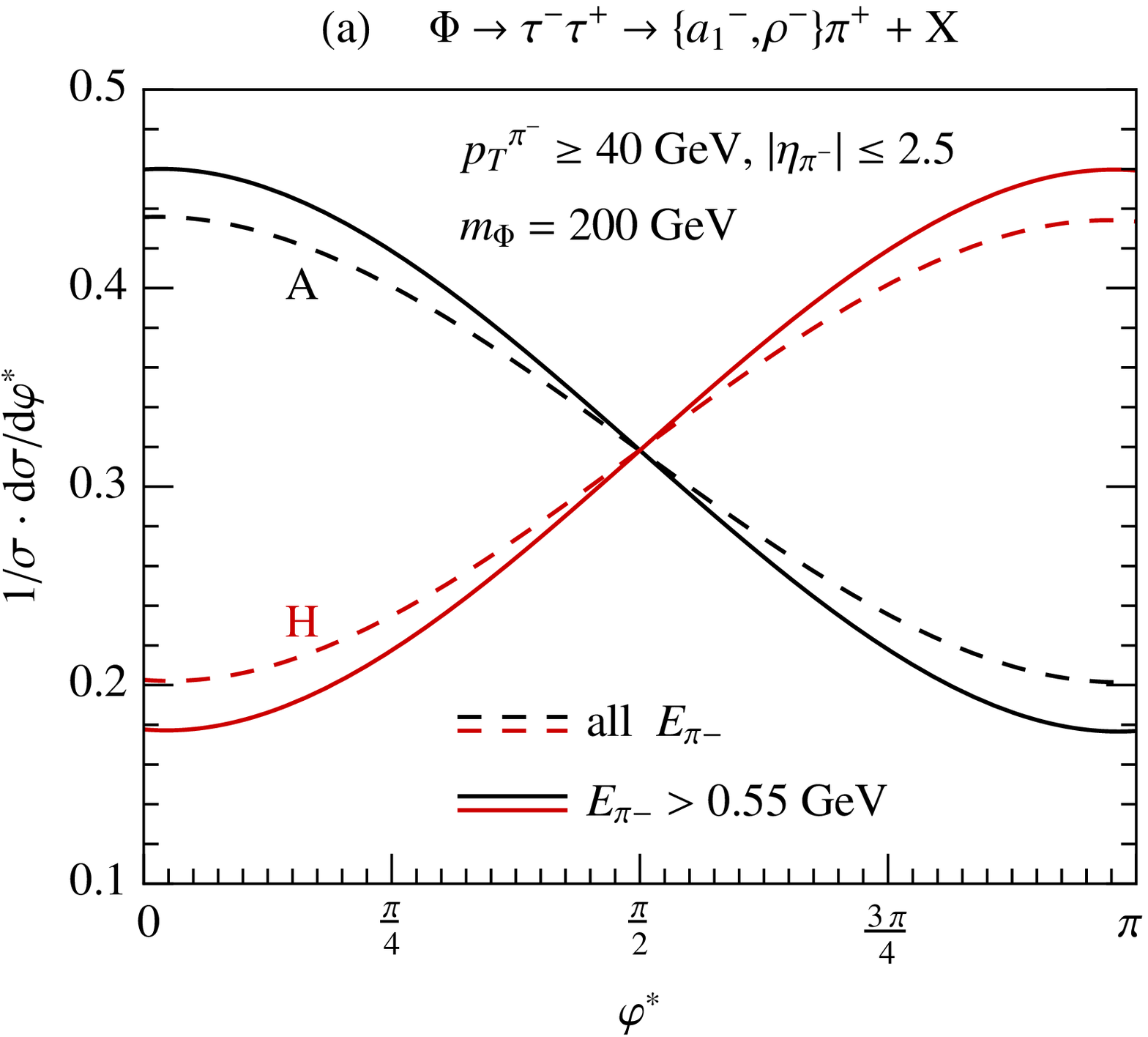}%
\hspace*{-.0cm}%
\includegraphics[scale=0.46]%
{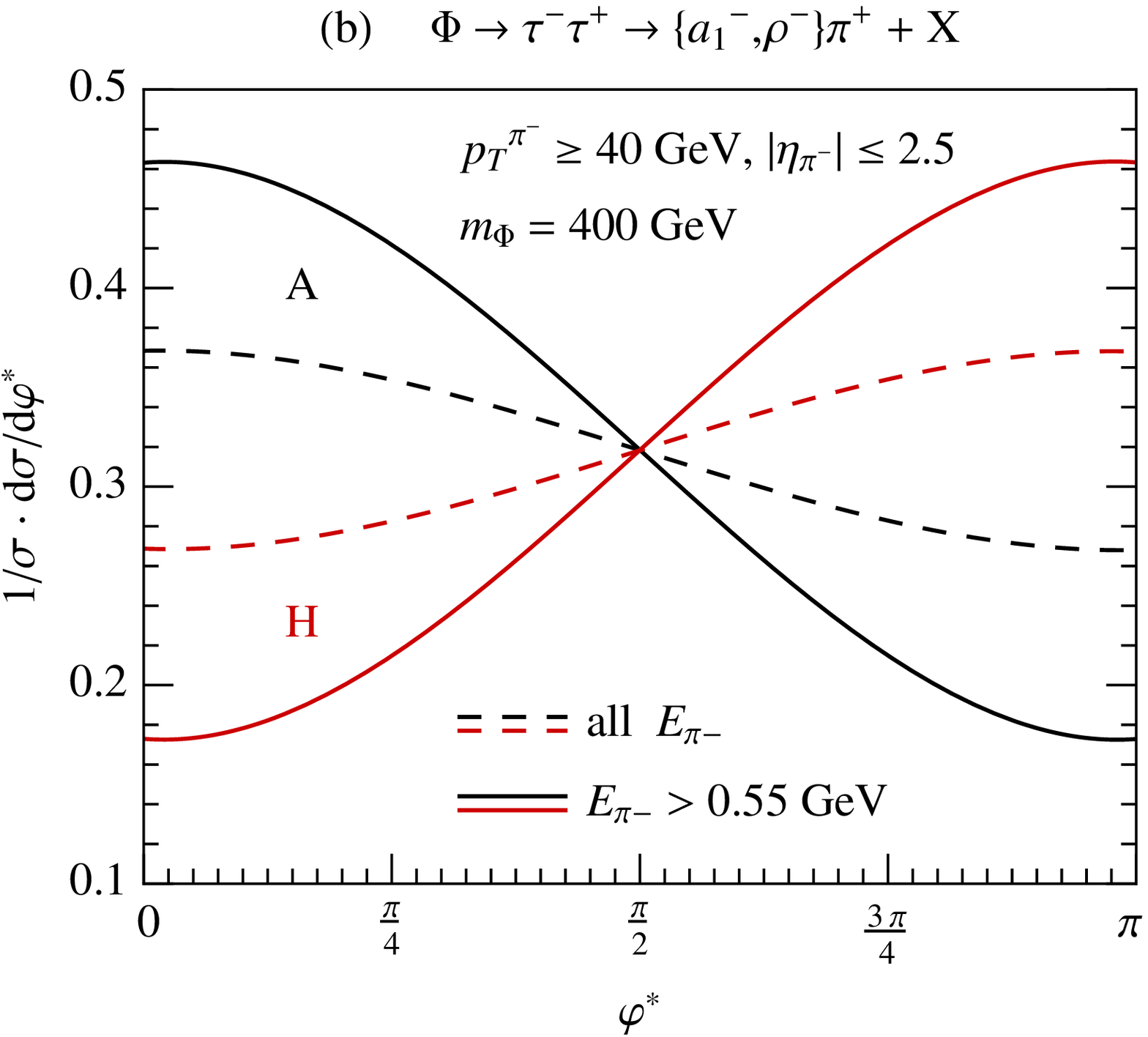}
\vspace*{-.4cm}
\caption{
  The normalized $\varphi^*$ distributions for the combined decays 
  $\Phi \to \tau^- \tau^+ \to \{a_{1}^-,\rho^-\} \pi^+ \to
  \pi^-\pi^+$.
  The $p_{T}$ and 
  $\eta$ cuts  (\ref{eq:Had_detector_cuts})
   are imposed on  $\pi^{-}$ only (dashed curves). The solid curves
   show  the distributions which result if instead of  (\ref{eq:Had_detector_cuts})
   an experimentally not feasible  cut
    on the pion energy $E_{\pi^{-}}$ in the $\tau$ rest frame 
    is applied. 
  (a) $m_{\Phi}=200$~GeV; 
  (b) $m_{\Phi}=400$~GeV.
\label{fig:rapi_phi_ptmi40_eta2.5_emitheo}
}
\end{figure}
As a consequence, the $\varphi^{*}$ distributions are dominated by 
contributions with positive values of the $\tau$-spin analyzer
functions $b_{\rho}(E_{\pi})$, $b_{a_1}(E_{\pi})$, and the 
distributions clearly differ for   $\Phi=H$ and $\Phi=A$, see 
Fig.~\ref{fig:rapi_phi_ptmi40_eta2.5_emitheo}(a).  
For small Higgs-boson masses, the cuts (\ref{eq:Had_detector_cuts})
are almost as  efficient as an experimentally not realizable  
cut on the pion energy in the $\tau$ rest frame, as shown by the 
solid curves  in Fig.~\ref{fig:rapi_phi_ptmi40_eta2.5_emitheo}(a). 
For heavy Higgs-bosons the discriminating power of the 
$\varphi^{*}$ distributions decreases, see
Fig.~\ref{fig:rapi_phi_ptmi40_eta2.5_emitheo}(b).
This decrease can be avoided by an additional cut, as discussed 
in Sec.~\ref{suse:recoapprox}.

In fact,  the situation is improved by taking into account 
the contribution from the direct decay $\tau^- \to \pi^- + 
\nu_{\tau}$ which has maximal $\tau$-spin analyzing power. This 
decay channel is less strongly affected by  the acceptance cuts 
(\ref{eq:Had_detector_cuts}) because  the $\pi^{-}$ energy in the 
$\tau^{-}$ rest frame is $E_{\pi^{-}} = m_{\tau}/2$. This is shown 
by the ratio $R$ of the contributions from the direct $\pi^-$ and 
the $\rho^- + a_1^-$ decays to the cross section $pp \to \Phi \to 
\tau^-\tau^+$, given for $m_{\Phi}=200$~GeV and $m_{\Phi}=400$~GeV 
in Table \ref{tab:sigma_decay_modes}. The numbers given in this table
were computed at tree-level, but we expect them to not be strongly
affected by radiative corrections to the $\Phi$ production amplitude.

\begin{table}[b]
\begin{tabular}{|c|c|c|c|c|}
\hline 
ratio \rule[-4mm]{0mm}{9mm}
    & ~$R_{no\, cuts}^{m_{\Phi}\,=\,200}$~ 
      & ~$R_{p_{T}^{\pi-},\eta_{\pi^{-}}\, cuts}^{m_{\Phi}\,=\,200}$~
        & ~$R_{no\, cuts}^{m_{\Phi}\,=\,400}$~
          & ~$R_{p_{T}^{\pi-},\eta_{\pi^{-}}\, cuts}^{m_{\Phi}\,=\,400}$~
\tabularnewline
\hline 
\rule[-3mm]{0mm}{7mm}
~$R = {\sigma_{\tau^-\to\pi^-}} / 
 {\sigma_{\tau^-\to\{a_{1}^-,\rho^-\}\to\pi^-}}$ 
  & $0.31$ & $0.82$ & $0.31$ & $0.50$
\tabularnewline
\hline
\end{tabular}
\caption{
  Ratio of different final-state contributions to the cross section 
  for $pp \to \Phi \to \tau^- \tau^+$ with and without detector cuts 
  as described in the text. 
\label{tab:sigma_decay_modes}
}
\end{table}

\begin{figure}[t]
\hspace*{-.15cm}
\includegraphics[scale=0.46]%
{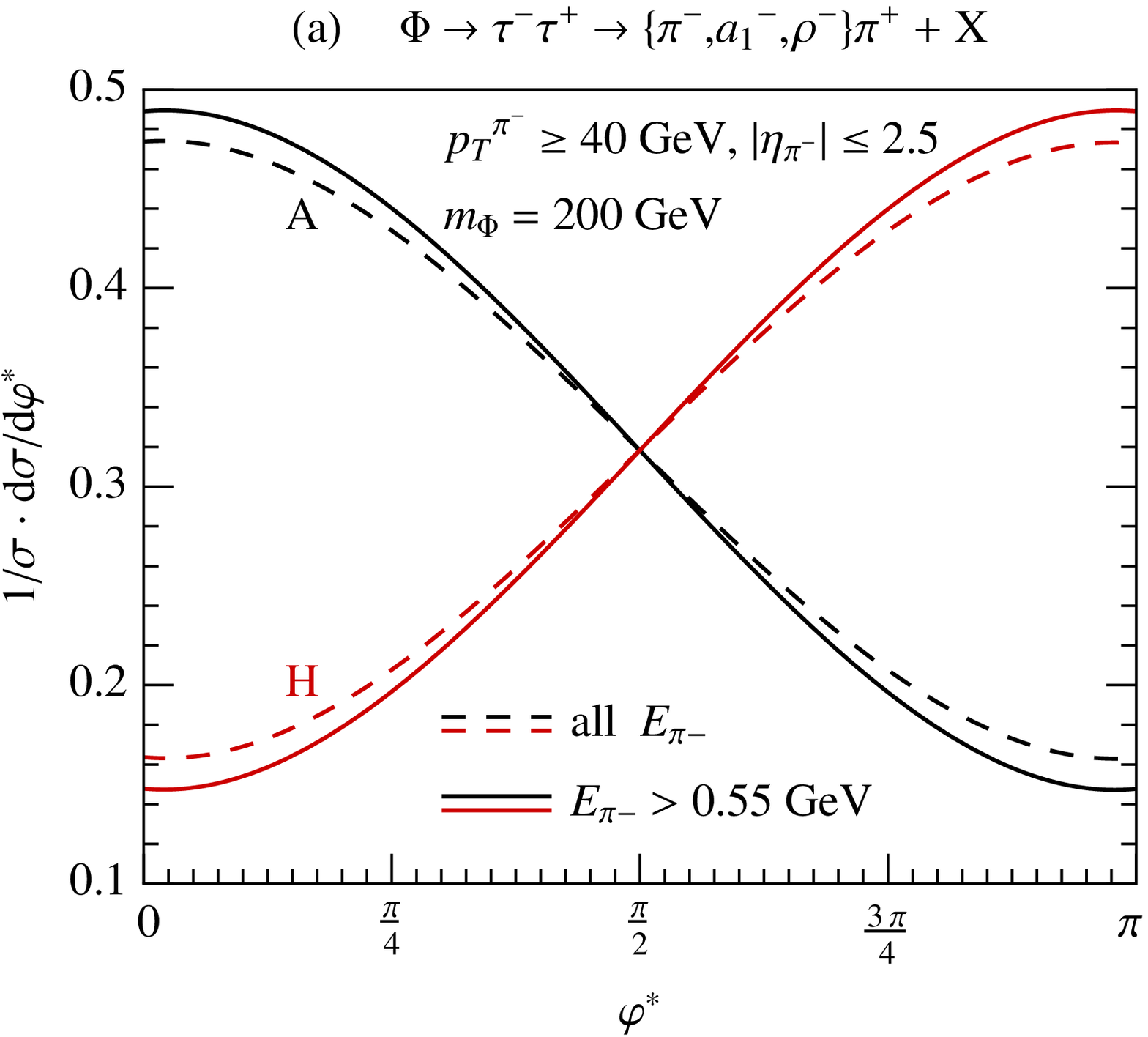}%
\hspace*{0.cm}%
\includegraphics[scale=0.46]%
{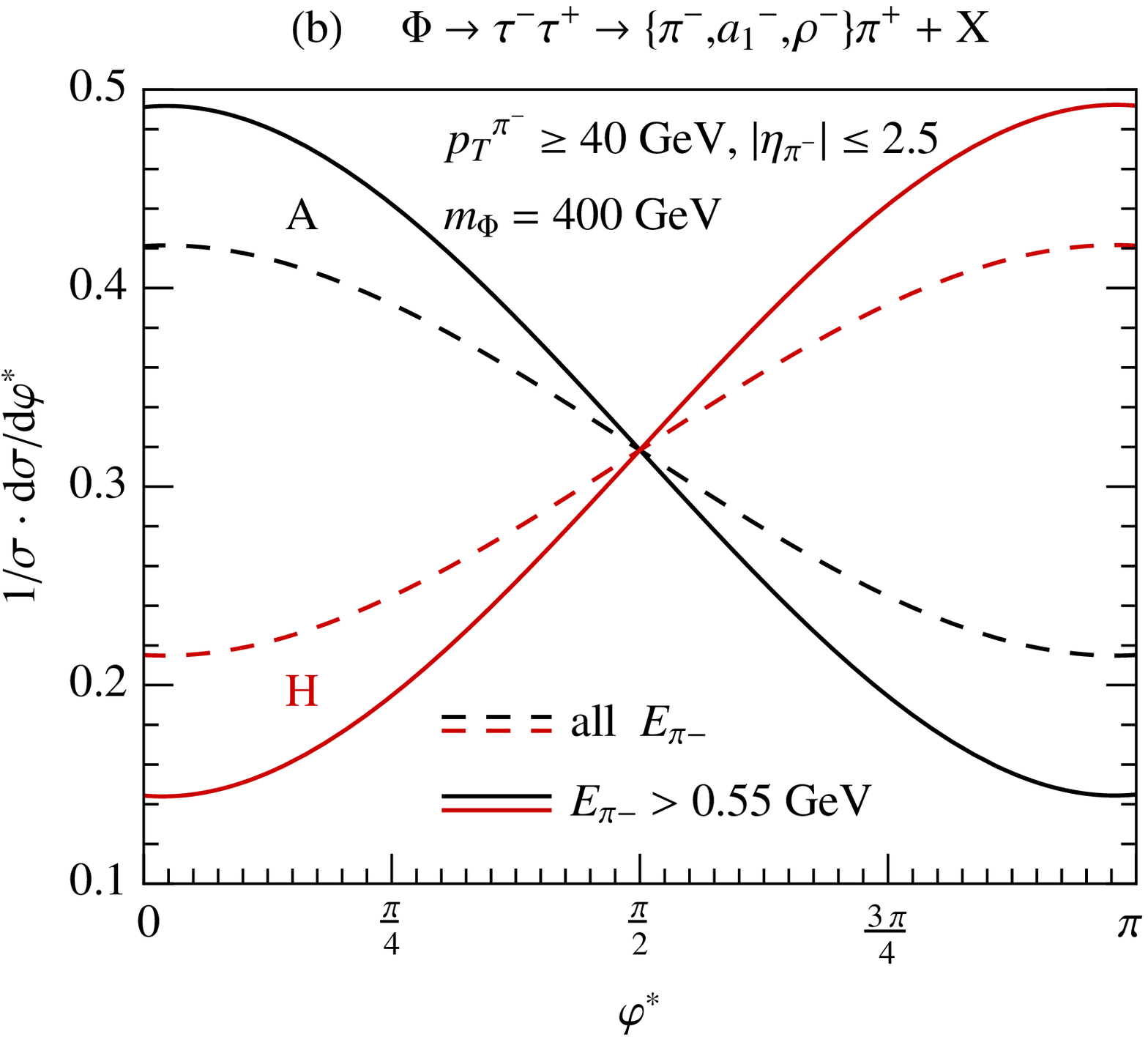}
\caption{
  Same as Fig.~\ref{fig:rapi_phi_ptmi40_eta2.5_emitheo}, including
  the contribution from the direct decay $\tau^{-} \to \pi^{-} \nu$.
  (a) $m_{\Phi}=200$~GeV; 
  (b) $m_{\Phi}=400$~GeV.
\label{fig:hadpi_cuts_mh200400_0.55Emi}
}
\end{figure}

The $\varphi^{*}$ distributions, with the direct $\tau^-\to \pi^-$
contribution included, are shown in 
Fig.~\ref{fig:hadpi_cuts_mh200400_0.55Emi}. Comparing with 
Fig.~\ref{fig:rapi_phi_ptmi40_eta2.5_emitheo} we see that the 
discriminating power  has indeed improved, both for small and 
large Higgs-boson masses. For $m_{H}=200$~GeV, the cuts 
(\ref{eq:Had_detector_cuts}) are in fact almost optimal, as 
one can see by comparing the dashed with the solid curves
in Fig.~\ref{fig:hadpi_cuts_mh200400_0.55Emi}.

The reconstruction of the $\varphi^{*}$ distributions requires
the  determination of the (normalized) impact parameter vectors 
${\bf n}_{-}$ and ${\bf n}_{+}$. One may ask whether 
a cut on their length would improve the sensitivity  of 
the data selected in this way. We have therefore performed a
simulation, along the lines outlined in \cite{Berge:2008dr}, where 
we require  $|{\bf n}_-| > 20\,\mu m$ for the 
displacement of the secondary vertex of $\tau^-\to \pi^-$, 
assuming an exponential decay of the $\tau$ with a mean 
life-time of $2.9 \cdot10^{-13}s$. However, it turns out  that 
the $\varphi^{*}$  distributions are only slightly affected -- there
is no gain in sensitivity. In addition, the cross section is reduced 
by almost a factor of 2. Therefore, we refrain from this requirement 
in the following. 


\subsection{Reconstruction of approximate $\tau$ momenta}
\label{suse:recoapprox}

From the discussion in the previous section we conclude that 
a more refined event selection is desirable for large Higgs-boson
masses of the order of $400\,\,{\rm GeV}$. In particular, 
additional cuts that remove events with low-energy pions (referred 
to the respective $\tau$ rest frame), 
where $b\left(E_{\pi}\right)$ is negative, would help to 
improve the discrimination  of $CP$-even and $CP$-odd resonances. 
Knowledge of the  $\tau^\mp$ 4-momenta  would allow us
to Lorentz-boost the measured pion momenta to the respective 
$\tau$ rest frame, where the application of a cut on $E_{\pi}$ 
would be straightforward. An approximate reconstruction of the $\tau$
momenta will be sufficient for this purpose, as long as it helps 
to enhance the difference of  $\varphi^{*}$ distributions for
scalar and pseudoscalar  Higgs-bosons. In the following we describe an 
approach where we  combine the information contained 
in the measured  pion momenta in the $pp$ laboratory frame  
and the experimentally known value of the Higgs-boson mass. 

\begin{figure}[t]
\hspace*{-.1cm}
\includegraphics[scale=0.45]%
{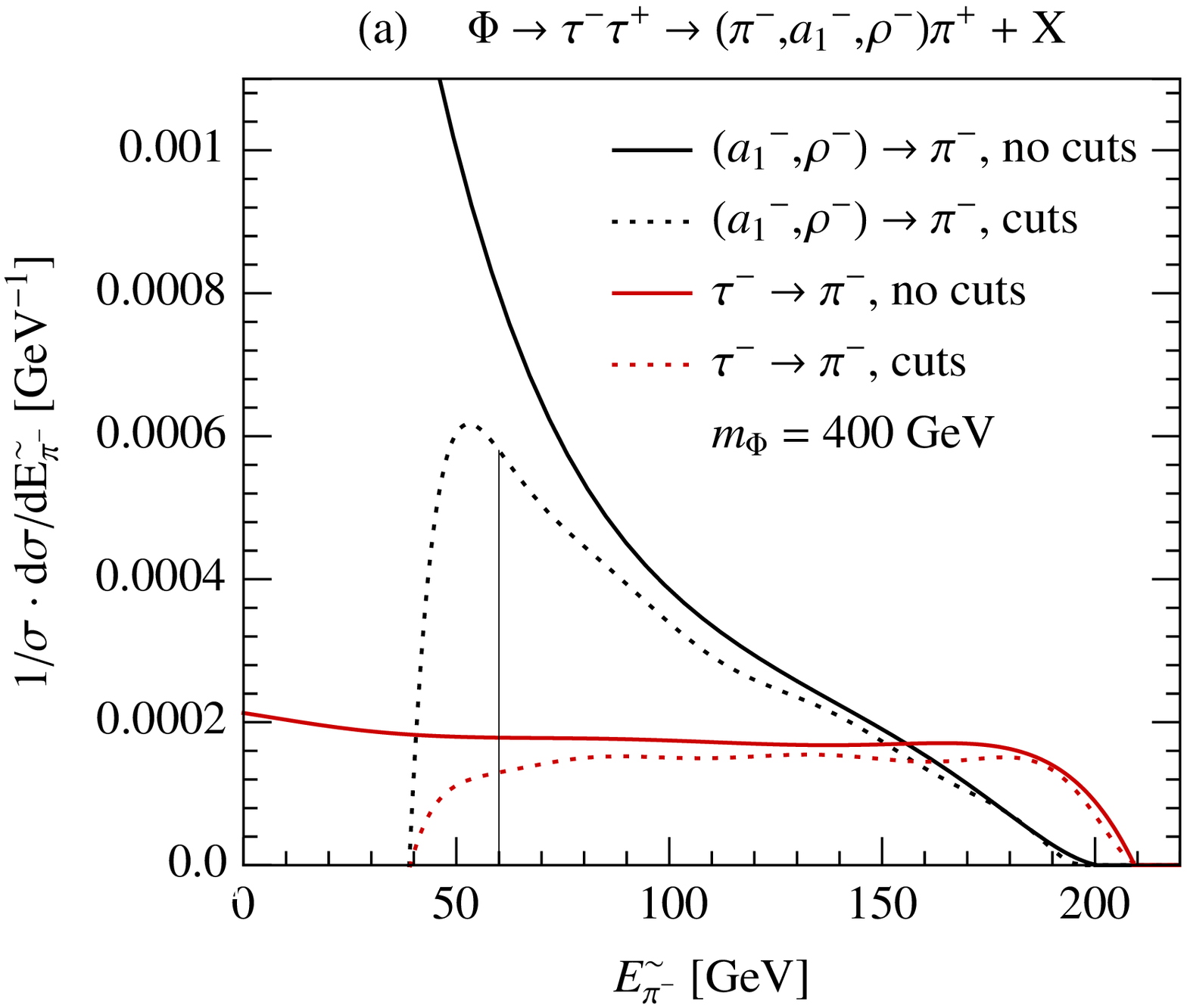}%
\hspace*{.2cm}
\includegraphics[scale=0.445]%
{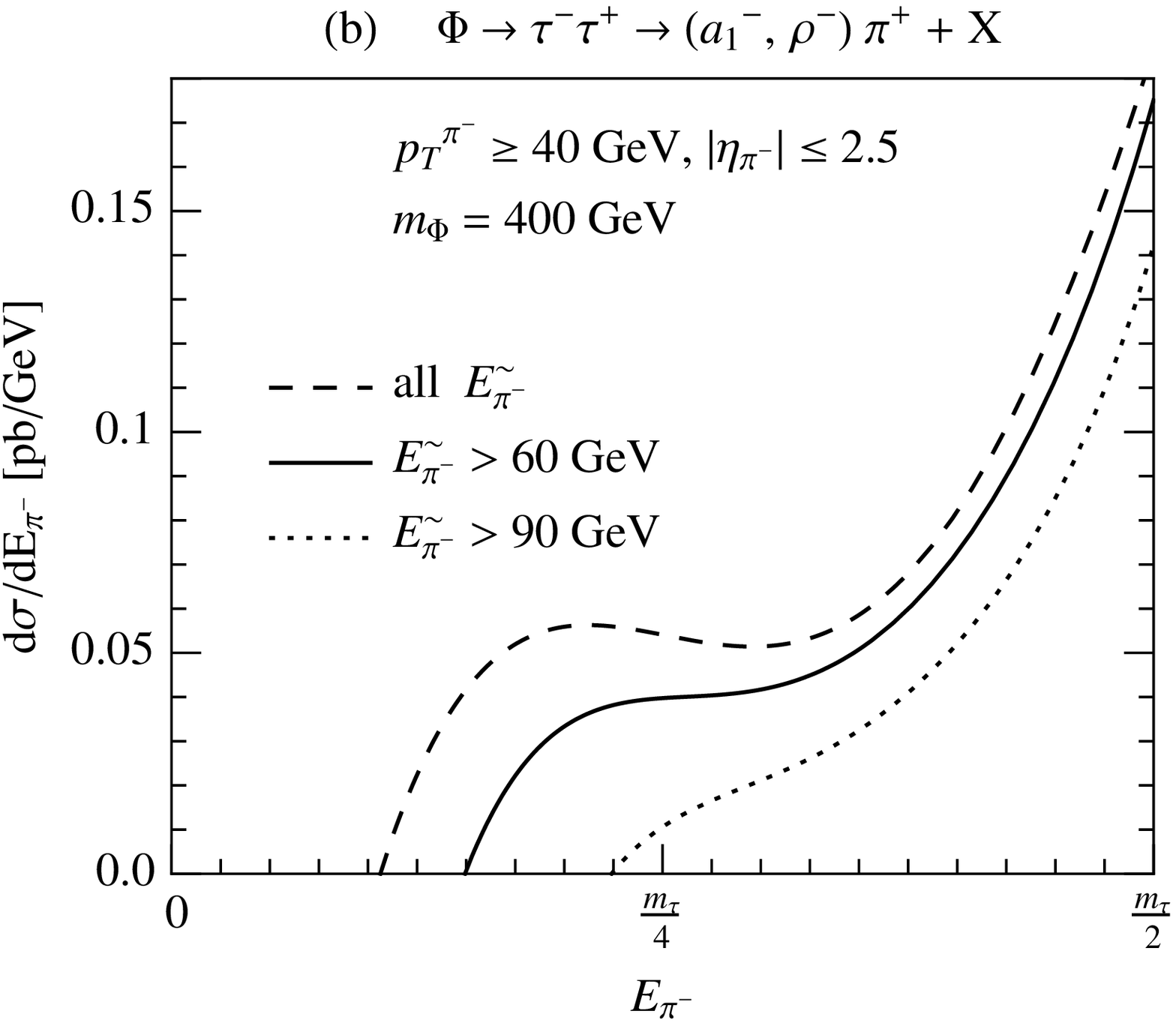}
\caption{
  (a) Distribution of $E^{\sim}_{\pi-}$ in the approximate 
  Higgs rest frame. The solid curves correspond to the distribution 
  without cuts; the dotted curves are obtained when applying 
  the cuts Eq.~(\ref{eq:Had_detector_cuts}). The vertical line
  indicates the additional cut $E^{\sim}_{\pi-}= 60$ GeV.
  (b) $E_{\pi}$ distribution  for the 
  combined decays  $\tau^-\to a_1^-, \rho^{-}$,   with cuts
  (\ref{eq:Had_detector_cuts}), with 
  and  without an additional cut on $E^{\sim}_{\pi-}$.
\label{fig:rapi_higgsapprox_60gev}
}
\end{figure}

We make the following approximations:
i) In the laboratory frame, the $\tau$ momenta ${\bf k}^{\pm}$ 
and $\pi$ momenta ${\bf p}_{\pm}$ are collinear, i.e.,  
${\bf k}^{\pm} = \kappa_{\pm} \hat{{\bf p}}^{\pm}$. 
ii)  The measured missing transverse momentum ${\bf P}^{miss}_{T}$ 
is assigned to the sum of the differences of the transverse
momenta of $\tau^\pm$ and the charged prong $a^\pm$, i.e.,
${\bf P}^{miss}_{T}  = {\bf k}^-_T - {\bf p}^-_T + {\bf k}^+_T - 
{\bf p}^+_T$.
This is a very crude approximation for  $\tau\to \rho, a_1$, but 
it serves the goal formulated above. One can then write down 
eight equations for the eight unknown components of the 
$\tau^{\pm}$ 4-momenta $k^{\pm \mu}$: 
\begin{eqnarray*}
p_{\Phi}^{\mu} & = & k^{+ \mu} + k^{- \mu} \, ,
\\
m_{\tau}^{2} & = & (k^{+})^2 \, ,
\\
m_{\tau}^{2} & = & (k^{-})^2 \, ,
\\
{\bf P}^{miss}_{T} & = & 
  {\bf k}^-_T - {\bf p}^-_T + {\bf k}^+_T - {\bf p}^+_T \, .
\end{eqnarray*}
These equations can be solved analytically. One obtains an 
approximate Higgs momentum $p_{\Phi}^{\sim\mu}$ which can be used
to boost the $\pi$ momenta to the corresponding approximate 
Higgs-boson rest frame. We denote the resulting pion energies 
by $E^{\sim}_{\pi^{-}}$. The distribution of $E^{\sim}_{\pi^{-}}$ 
is shown in Fig.~\ref{fig:rapi_higgsapprox_60gev}(a).
The upper solid (black) curve shows the distribution for the 
$\rho+a_1$ decays without cuts, while the dotted curves result
from imposing the cuts (\ref{eq:Had_detector_cuts}). The lower 
solid and dotted (red) curves display the corresponding 
distribution for the direct $\tau^- \to \pi^- \nu$ decay. From 
this result and the analysis of  the previous section we conclude 
that events with small pion energies in the $\tau$ rest frames 
are correlated with events with small pion energies in the 
approximate Higgs-boson rest frame. This statement is supported 
by the distribution displayed  in 
Fig.~\ref{fig:rapi_higgsapprox_60gev}(b). This figure shows 
that, by imposing a cut on $E^{\sim}_{\pi^{-}}$, the distribution 
of $E_{\pi^{-}}$ is shifted to larger values of the pion energy. 

\begin{figure}[t]
\hspace*{0cm}%
\includegraphics[scale=0.46]%
{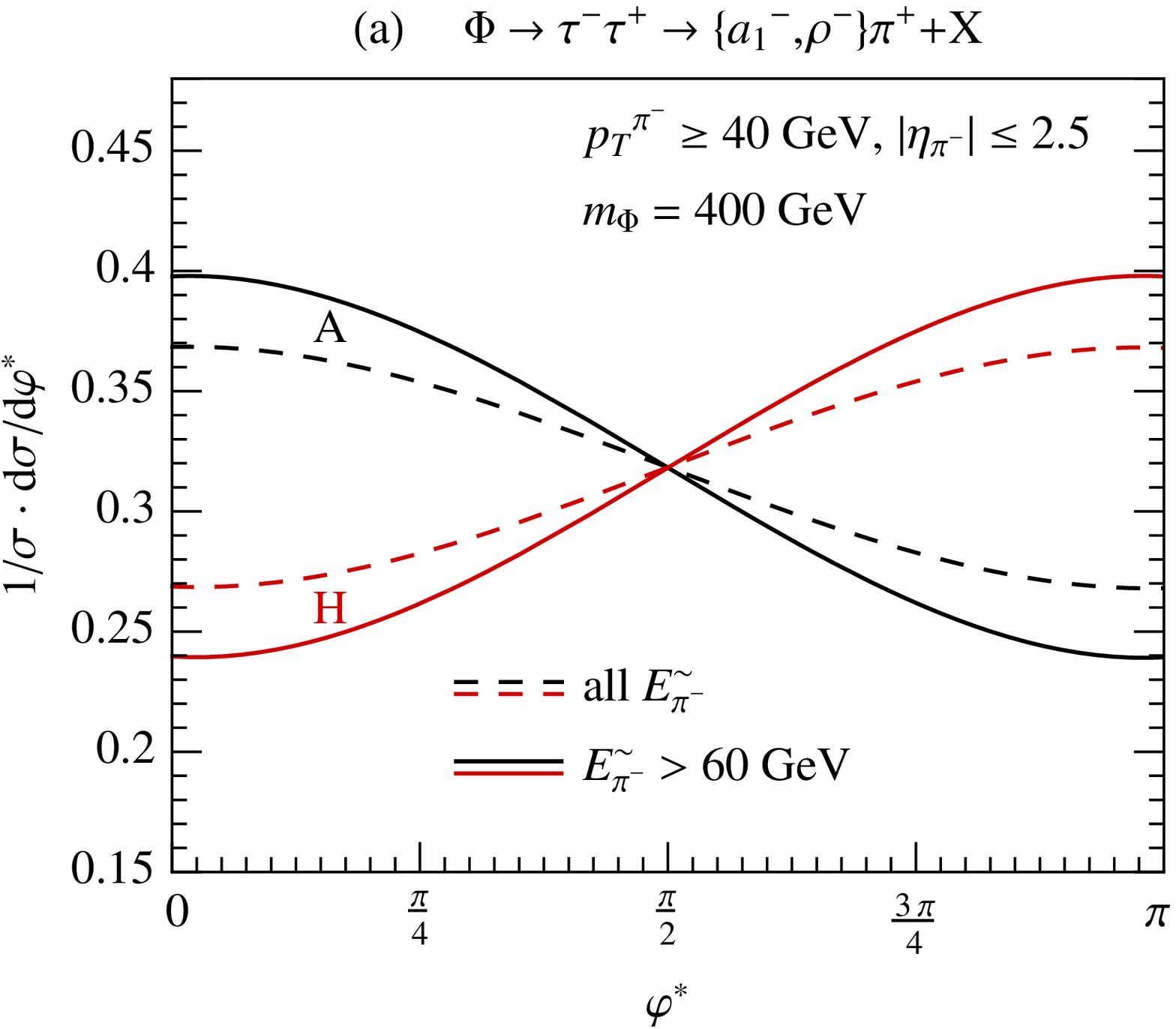}%
\includegraphics[scale=0.46]%
{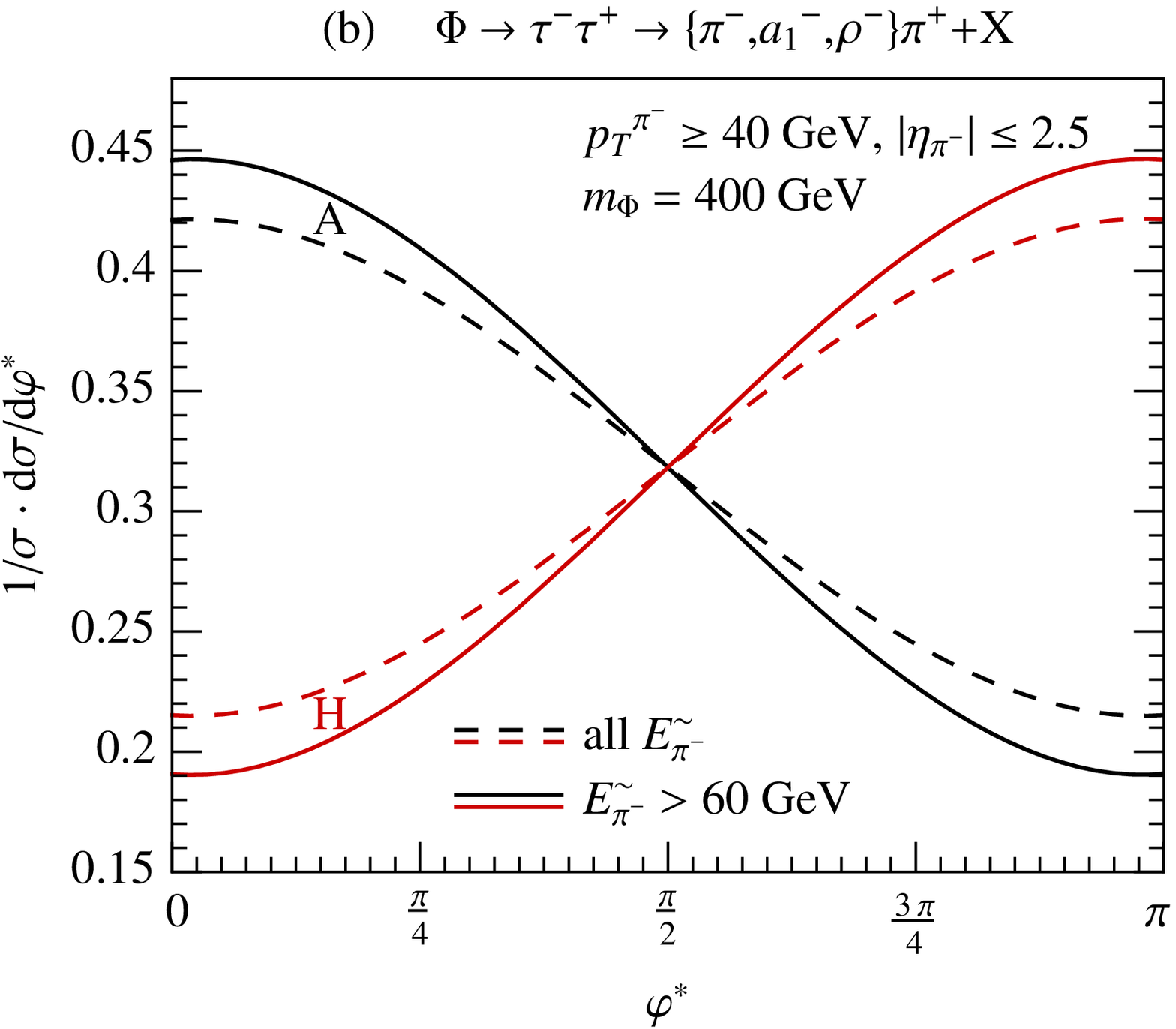}
\caption{
  (a) The normalized $\varphi^{*}$ distributions (without the 
  direct $\tau^{-}\to\pi^{-}$ decay) with and without a cut on 
  the reconstructed pion energy $E^{\sim}_{\pi-}$. The solid 
  curves result from events  with  $E^{\sim}_{\pi^{-}} > 60$~GeV. 
  (b) Same as (a), but with the direct $\tau^{-} \to
  \pi^{-}$ decay channel included. 
\label{fig:hadpi_higgsapprox_60gev}
}
\end{figure}

Encouraged by these observations we calculate, for heavy 
Higgs-bosons, the normalized $\varphi^{*}$ distributions by imposing 
the additional cut $E^{\sim}_{\pi^{-}} > 60$~GeV. The results 
are shown in Fig.~\ref{fig:hadpi_higgsapprox_60gev}. 
For comparison, the dashed curves  result from applying only the 
cuts (\ref{eq:Had_detector_cuts}).  The additional cut of 
$E^{\sim}_{\pi^{-}} > 60$~GeV clearly leads to  an increase
in sensitivity. Including the $\tau \to \pi \nu$ decay leads 
to a further improvement, see 
Fig.~\ref{fig:hadpi_higgsapprox_60gev}(b). In fact, the additional 
cut affects the direct decay channel only marginally. 

One should keep in mind that the additional cut on $E^{\sim}_{\pi^-}$ 
will reduce the size of the event samples for the measurement
of the $\varphi^{*}$ distributions. The sample based on
$\tau^-\to \rho^-, a_1^-$, and $\pi^-$ decays will  be reduced by about   
$18\,\%$. An analysis  including a full detector simulation is 
required in order to optimize the selection  cuts
(\ref{eq:Had_detector_cuts}) and the cut on $E^{\sim}_{\pi}$.  

These conclusions will not be affected by analyzing different 
$\Phi$ production processes or by taking into account 
higher-order QCD corrections. For example, let us consider
the production of a  Higgs-boson  with very  high $p_T$ in the
laboratory frame. Then the $l^\pm$ and $\pi^\pm $ from
$\tau^\pm$ decays can  pass the transverse momentum cuts 
(\ref{eq:lepton_detector_cuts}) and (\ref{eq:Had_detector_cuts}), 
even if the energies of the charged prongs in the respective 
$\tau$ rest frames are small. As a result, contributions to the 
decay modes $\tau \to l,a_1,\rho$ with $b(E_{\pi,l}) >0 $ and 
$b(E_{\pi,l}) <0 $ cancel and the discriminating power of the
$\varphi^*$ distribution is reduced. However, when applying an  
additional cut in the approximate Higgs rest frame as described 
above, the dangerous contributions will be rejected. This yields 
basically  the same  $\varphi^*$ distributions as before. 
In addition, for Higgs events with large $p_T$, much better 
methods of reconstructing the $\tau$ rest frame can be applied, 
for instance the collinear approximation~\cite{Ellis:1987xu}
or the method described in~\cite{Elagin:2010aw}. High $p_T$ 
particles/jets that are produced in association with a 
Higgs-boson can also be used to reconstruct an approximate Higgs rest 
frame.


\subsection{Combined leptonic and hadronic 1-prong decays}
\label{suse:combleha}

In this section we present results for the $\varphi^{*}$ distributions
taking into account all 1-prong decays (\ref{eq:tau_decay_channels}),
i.e., 
\vspace*{-3pt}
\begin{equation}
\label{all1pkanal}
pp \to \Phi \to \tau^- \tau^+ \to \left\{
	    \begin{array}{l}
		l^- \ l'^+ + X, \qquad  l, l' = e, \mu , \\
		l^- \pi^+ + X  \quad {\rm and} \quad \pi^- l^+ + X, \\
		\pi^- \pi^+ + X  \, .  \\
	    \end{array}  \right. \\[2pt]
\end{equation}
We refer to the different channels in (\ref{all1pkanal}) by
{\it dilepton}, {\it lepton-pion}, and {\it two-pion} final 
states. The cuts (\ref{eq:lepton_detector_cuts}) and 
(\ref{eq:Had_detector_cuts}) are applied to the charged 
leptons and pions, $l^\mp$ and $\pi^\mp$, respectively.

\begin{figure}[t]
\hspace*{0cm}%
\includegraphics[scale=0.46]%
{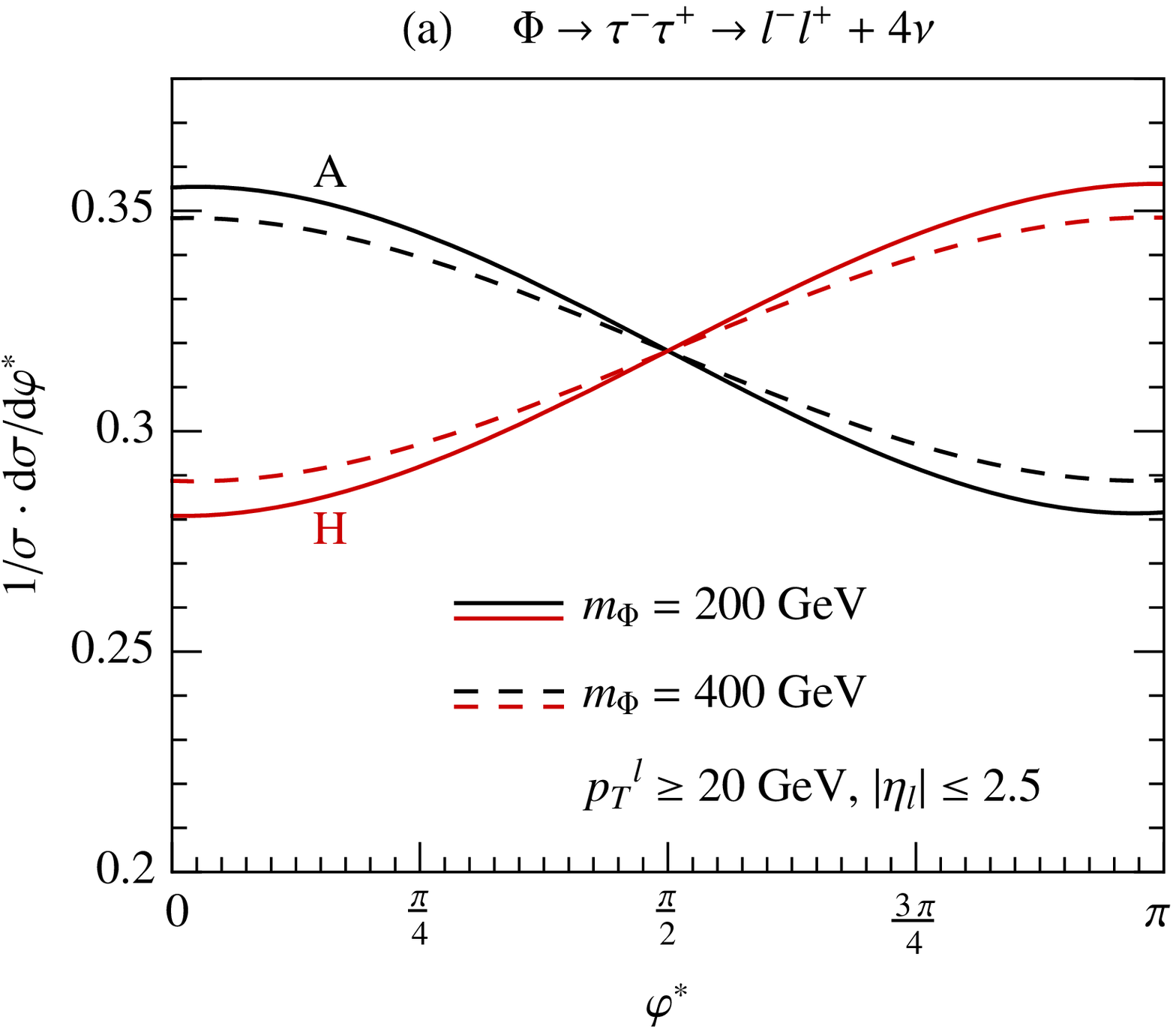}%
\includegraphics[scale=0.46]%
{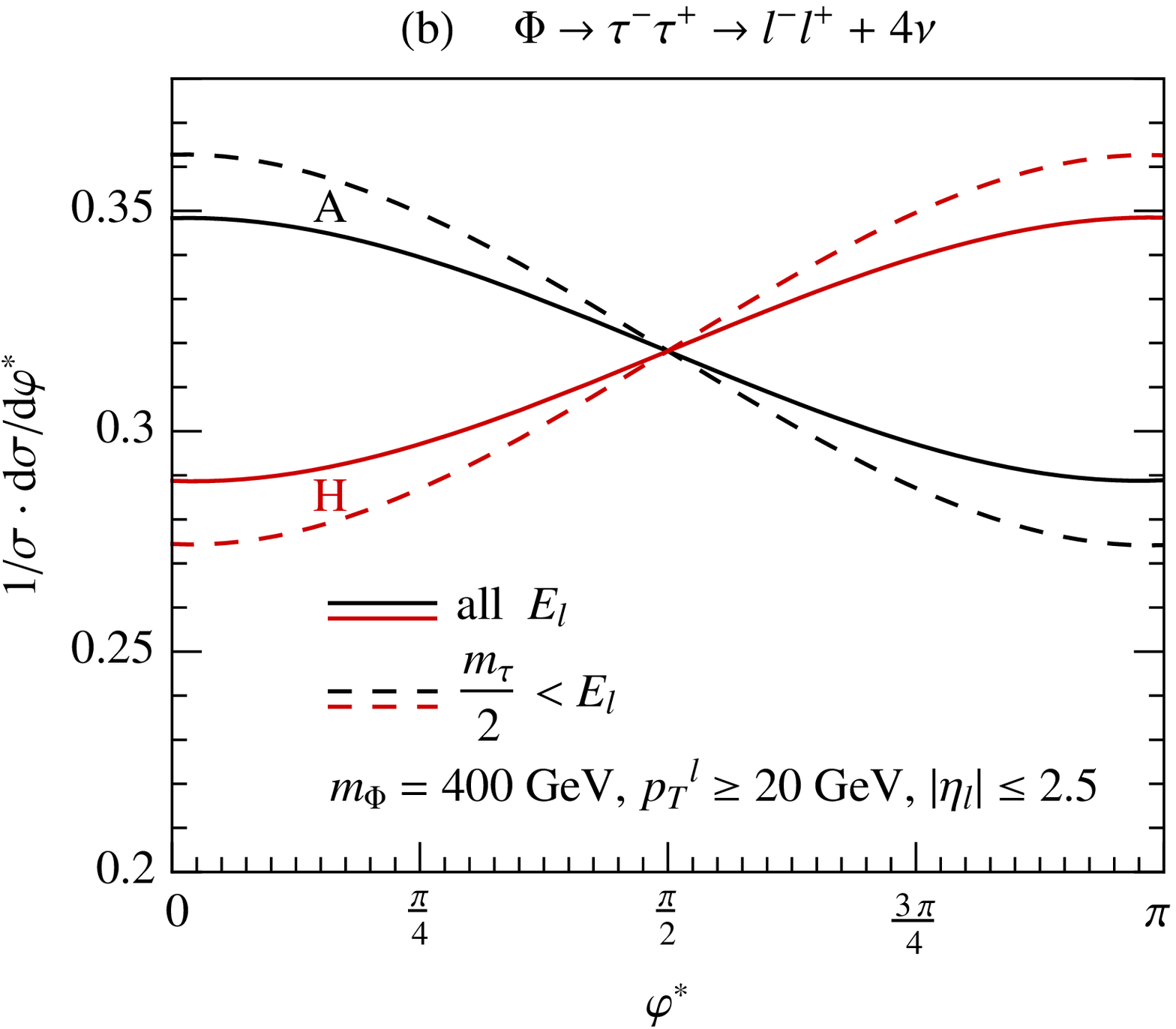}
\caption{ (a)
  The normalized $\varphi*$ distributions for the dilepton final states, for
   $m_{\Phi} = 200$~GeV (solid curves) and 
  $m_{\Phi} = 400$~GeV (dashed curves). 
  (b) The solid curves are identical to  (a) for $m_{\Phi} = 400$~GeV;
  the dashed curves result from applying  an additional cut 
  $E_{l} > \frac{m_{\tau}}{2}$ on the charged lepton energies in the
  $\tau^\mp$ rest frames.
\label{fig:leplep_phi_ptmi20_eta2.5}
}
\end{figure}

For the dilepton final states the $\varphi*$ distributions 
are presented in Fig.\ \ref{fig:leplep_phi_ptmi20_eta2.5}(a) 
for two values of the Higgs-boson mass. The figure shows that 
the power  of   $\varphi*$ to discriminate between a scalar and 
pseudoscalar Higgs-boson is, in these decay channels, almost 
independent of the mass of $\Phi$. As discussed in 
Sec.~\ref{suse:leppion}, one would increase the sensitivity 
if one could reconstruct the $\tau^\mp$ rest frames and select 
an event sample with  an additional  cut on the lepton energies 
in these frames. However, Fig.~\ref{fig:leplep_phi_ptmi20_eta2.5}(b) 
shows that the enhancement would be  rather modest.

\begin{figure}[t]
\hspace*{-.05cm}%
\includegraphics[scale=0.46]%
{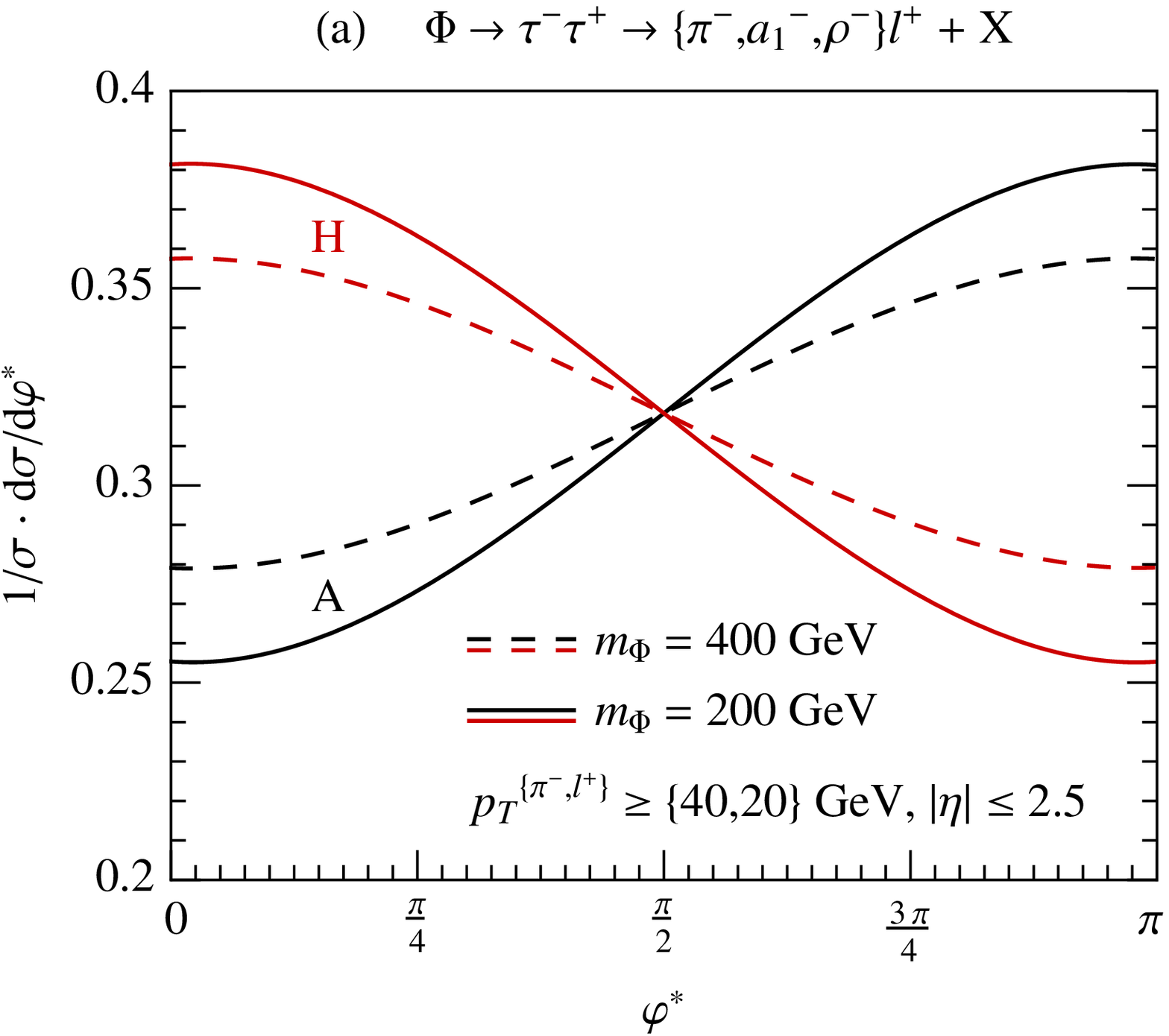}%
\hspace*{.1cm}%
\includegraphics[scale=0.46]%
{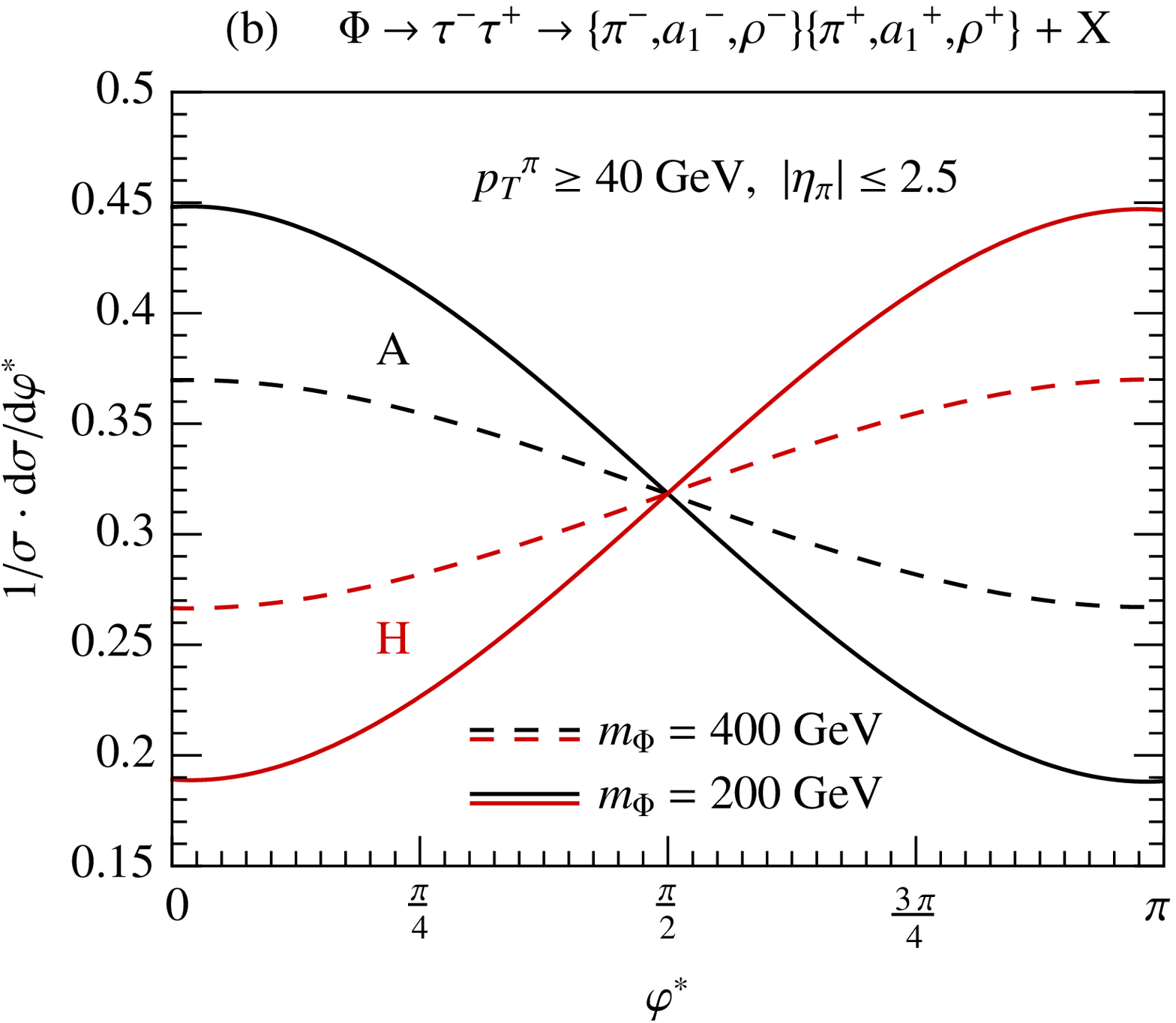}
\caption{
  (a) The  normalized $\varphi*$ distributions  for the  lepton-pion 
  final states for $m_{\Phi} = 200$~GeV (solid curves) and $m_{\Phi} = 
  400$~GeV (dashed curves). 
  (b) The  $\varphi*$ distributions for the two-pion final states.
\label{fig:hadlep_phi_pt4020_eta2.5}
}
\end{figure}

In Fig.~\ref{fig:hadlep_phi_pt4020_eta2.5}(a) and (b) the 
$\varphi^{*}$ distributions are presented for the lepton-pion 
and two-pion final states, respectively, for two different values 
of the Higgs-boson mass. For these final states, the discriminating 
power of the $\varphi^{*}$ distribution decreases with increasing 
Higgs-boson mass. We emphasize, however, that our evaluation is 
conservative in the sense that we applied only the  acceptance cuts
(\ref{eq:lepton_detector_cuts}) and (\ref{eq:Had_detector_cuts}).
As shown in Sec.~\ref{suse:recoapprox}, a  further cut  on the
charged-pion energy $E^{\sim}_{\pi}$ in the approximate Higgs-boson
rest frame would significantly enhance the discriminating power of
$\varphi^{*}$ in the case of heavy Higgs-bosons or Higgs-bosons with
large $p_T$.

Notice  that the $\varphi^{*}$ distributions for a scalar 
(pseudoscalar) Higgs-boson have opposite slopes for  lepton-pion 
and two-pion final states. This is due to the fact that the signs
of the leptonic and hadronic spin analyzer functions 
$b(E_{l})$ and $b(E_{\pi})$ differ, both in the  low-energy and
high-energy part of the spectrum.  Therefore, a  very good
experimental  discrimination of leptons and pions will be crucial 
for this measurement. 

It should be noticed that i) a Higgs-boson with scalar and pseudoscalar
$\tau$-Yukawa couplings of equal strength (i.e., an ideal $CP$ mixture)
or ii) (nearly) mass-degenerate scalar and pseudoscalar Higgs-bosons
with equal production cross sections yield a $\varphi^*$ distribution
which is flat, both for dilepton, lepton-pion, and two-pion final
states (cf.\ \cite{Berge:2008dr}). In order to unravel these
possibilities, one has to measure  the distribution of the angle
$\psi^{*}_{CP}$, see below. The $\varphi^*$ distributions of 
mass-degenerate scalars and pseudoscalars with  different reaction 
cross sections and of a $CP$ mixture with $|a_{\tau}|\neq |b_{\tau}|$  
have shapes which lie between the pure scalar and
pseudoscalar cases and can also be disentangled  with a joint
measurement of the  $\psi^{*}_{CP}$ distribution. 
 
Next, we make a crude estimate of how many events are needed in 
order to distinguish between a scalar and a pseudoscalar
Higgs-boson in the different decay channels (\ref{all1pkanal}). 
We consider the asymmetry 
\begin{equation}
\label{Aphiasy}
A_{\varphi^*} =  
\frac{N(\varphi^*>\pi/2) - N(\varphi^*<\pi/2)}%
{N(\varphi^*>\pi/2) + N(\varphi^*<\pi/2)} \, .
\end{equation}
The asymmetries can be computed for the different final states 
from the distributions Figs.~\ref{fig:leplep_phi_ptmi20_eta2.5}(a), 
\ref{fig:hadlep_phi_pt4020_eta2.5}(a),  and (b), for a scalar and 
a pseudoscalar Higgs-boson. Assuming that systematic effects can 
be neglected, we estimate from these asymmetries the event numbers 
needed to distinguish a scalar from a pseudoscalar Higgs-boson with 
3 standard
deviation (s.d.) significance. These numbers are given in 
Table~\ref{tab:eventnrphi}.

\begin{table}[t]
\begin{center}
\begin{tabular}{|c|c|c|c|}
\hline 
~$m_\Phi$ [GeV]~
    & ~dilepton~
      & ~lepton-pion~
        & ~two-pion~
\tabularnewline
\hline 
200 & 380 & 116 & 18
\tabularnewline
400 & 600 & 334 & 207
\tabularnewline
\hline
\end{tabular}
\end{center}
\caption{Event numbers needed to distinguish between a $CP$-even and
 $CP$-odd spin-zero state $\Phi$ with 3 s.d.\ significance.
 \label{tab:eventnrphi}
}
\end{table}

One may ask how 
vulnerable the $\varphi^*$ distributions -- and the $\psi^{*}_{CP}$  
distributions given in the next section -- are with respect to 
uncertainties in the experimental determination of the $\Phi$ 
production/decay vertex and of the energies and momenta of the 
charged prongs. This question was investigated in \cite{Berge:2008dr} 
for the direct pion decay modes $\tau^-\tau^+\to \pi^-\pi^+ 
{\bar\nu}_{\tau} \nu_{\tau}$ by a Monte Carlo simulation and it 
was found that these distributions retain their discriminating
power when measurement errors are taken into account. One may assume
that this result stays valid also for the larger class of 1-prong 
decay modes considered in
this paper. 

We close this section with a brief discussion of background reactions
 to the  $\Phi \to \tau^-\tau^+$ signal. These include QCD multijets,
 $t{\bar t}$, single top, $W$ + jets, $Z/\gamma^*$ + jets, $WW$, $WZ$,
 and $ZZ$ production. Among these,  $Z/\gamma^* \to \tau^-\tau^+$ (+
 jets) constitutes an essentially irreducible background for a  
Higgs-boson with mass close to the $Z$ mass. 
 Most of this background can be distinguished from the signal by means
 of appropriate discriminating variables, in particular 
 by reconstructing the $\tau$-pair invariant mass 
 $M_{\tau\tau}$  using a likelihood
 technique, with which a mass resolution of
 $\sim 21\%$ was achieved~\cite{Chatrchyan:2011nx}. 
  For Higgs-bosons with masses $m_{\Phi}\gtrsim 200$ GeV,
  the $Z^*\to \tau^-\tau^+$ background can be suppressed 
   by appropriate cuts on $M_{\tau\tau}$, which works also for purely
   hadronic $\tau^-\tau^+$ decays~\cite{Perchalla2011}.

  If a Higgs-boson will be
  found with a mass not too far away from the $Z$ mass one may, 
   in the long run of the LHC, resort to the production of
   $\Phi$ by vector boson fusion, in order to study the above
   distributions. In vector boson fusion, $\Phi$ is produced in the
   central region, which provides a good veto against QCD background.
  More importantly, one has an additional signature from two 
  well-separated forward jets, which gives a veto against $Z^*\to
  \tau^-\tau^+$. Yet, in the SM and for large portions of the parameter
  spaces of models with an extended Higgs sector, 
  $gg \to \Phi$ (and $b {\bar b} \to \Phi$ for
  large $\tan\beta$) is by far the dominant $\Phi$ production process.
  The modulus of the
   pseudorapidity, $|\eta|,$
  of  a light Higgs-boson produced in these reactions is large, while
   its transverse momentum, generated by  QCD radiation, is
   small on average. The $\tau \tau$ pair from $\Phi\to \tau^-\tau^+$
   is balanced in its total transverse momentum and its sum of the azimuthal
   angles. This provides a good discrimination against the QCD
   background, but not against $Z\to \tau^-\tau^+$. At this point one
  may exploit spin effects. The chiral-invariant $Z\tau\tau$ and
    $\gamma\tau\tau$ couplings
  lead to characteristic $\tau\tau$ spin correlations (see, for
  instance,
  \cite{Bernreuther:1989kc,Bernreuther:1993nd,Pierzchala:2001gc,Rosendahl:2011si})
  which differ from those that result from the decay of a spin-zero
  resonance whose fermion couplings are chirality-flipping. 
  For instance, for the
 decays $\tau^-\tau^+ \to \pi^-\pi^+\nu {\bar \nu}$, this has the
 following consequence. If the  $\pi^-\pi^+$ result from
 $Z$ boson decay, the number of  $\pi^-\pi^+$ events with
 $E_{\pi^-}$ and  $E_{\pi^+}$ both large  or both small (in the 
  $\tau^-\tau^+$ ZMF) is much larger than the number of events
 with $E_{\pi^-}$ large (small) and  $E_{\pi^+}$ small (large); while
  for $\Phi \to \tau^-\tau^+$ just the opposite is the case. These $\pi\pi$
  energy distributions may be used to discriminate the signal from
   the irreducible background in the case of a light $\Phi$.
   In addition, also the  $\pi^-\pi^+$ invariant mass distribution
   shows some difference between events from  $\Phi \to \tau^-\tau^+$ and
   $Z \to \tau^-\tau^+$ \cite{Pierzchala:2001gc}.
     Rather than trying to discriminate
   against the irreducible background, an alternative strategy might
   be, for a light $\Phi$,    to take into account
 the $Z\to \tau^-\tau^+$ events 
   both in the measurement and in the Monte Carlo
 modeling  of the 
 distributions (\ref{phistar}) and (\ref{eq:psi_star_distribution}).
 This requires a detailed study which is, however, beyond the scope of
 this paper.


\subsection{Higgs-sector $CP$ violation}
\label{suse:Hscp}

Besides $\varphi^*$, a further important observable in this context 
is the angle $\psi^{*}_{CP}$ defined in 
(\ref{eq:psi_star_distribution}).
It is the appropriate variable to check whether or not a spin-zero 
resonance $\Phi$ has couplings to both scalar and pseudoscalar 
$\tau$ lepton currents. A nontrivial $\psi^{*}_{CP}$
distribution, respectively a nonzero asymmetry associated
with this  distribution would be evidence for $CP$ violation in the
``Higgs sector'' (which is different from  Kobayashi-Maskawa $CP$ 
violation).  Such a discovery would have enormous consequences, 
in particular for baryogenesis scenarios (see, e.g., the reviews 
\cite{Cohen:1993nk,Bernreuther:2002uj}).  

We assume here that $\Phi$ is an ideal mixture of a $CP$-even and
$CP$-odd spin-zero state, with reduced Yukawa couplings $a_{\tau} = 
- b_{\tau}$ to $\tau$ leptons\footnote{Suffice it to mention that 
for the normalized $\psi^{*}_{CP}$ distribution only the relative 
magnitude and phase of $a_{\tau}$ and  $b_{\tau}$ matter, while the 
magnitudes of these couplings determine the decay rate of $\Phi\to 
\tau\tau$.}. For definiteness, we take  $a_{\tau} = - b_{\tau}=1$.
We call this the {\it CPmix} scenario for short and consider it for 
a Higgs-boson with mass $m_{\Phi}=200$ and $400$ GeV. For 
comparison we consider also three scenarios where $CP$ is conserved: 
i) a pure scalar $H$, ii) a pure pseudoscalar $A$, and iii) the case 
of a (nearly) mass-degenerate scalar $H$ and pseudoscalar $A$ with 
approximately the same production cross section and decay rate 
into $\tau$ leptons.

\begin{figure}[t]
\hspace*{-.15cm}
\includegraphics[scale=0.47]{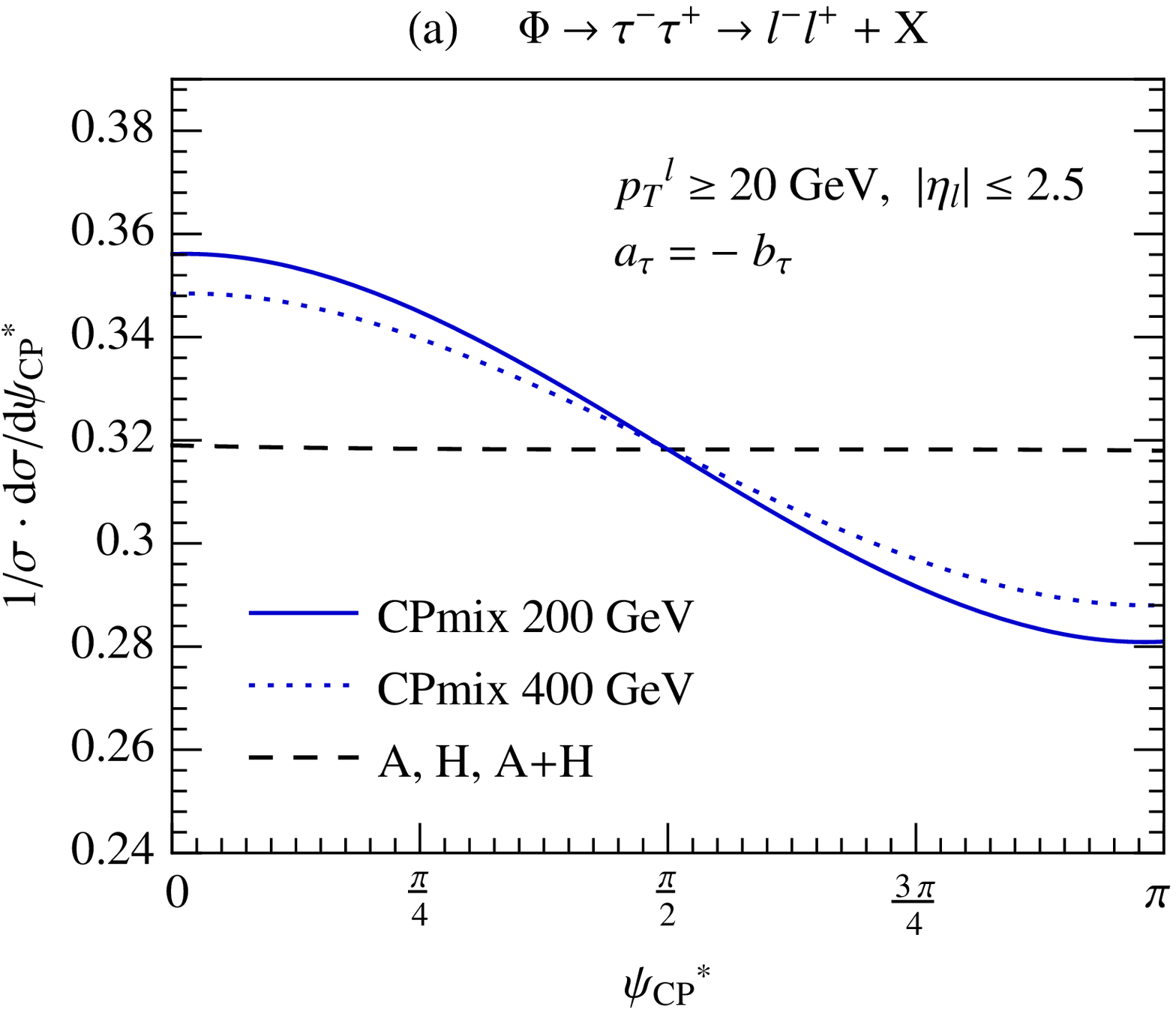}%
\includegraphics[scale=0.47]{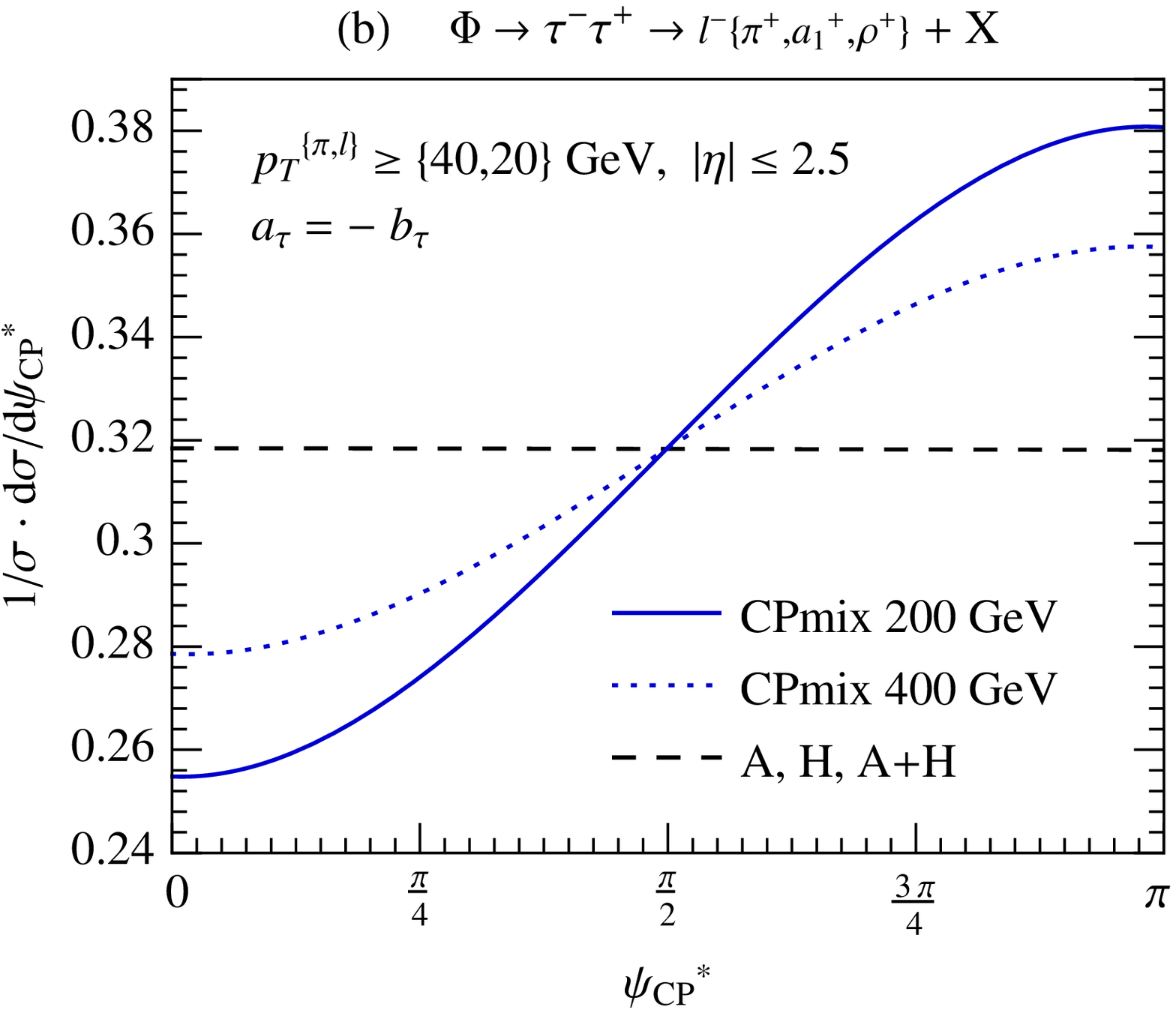}
\includegraphics[scale=0.47]{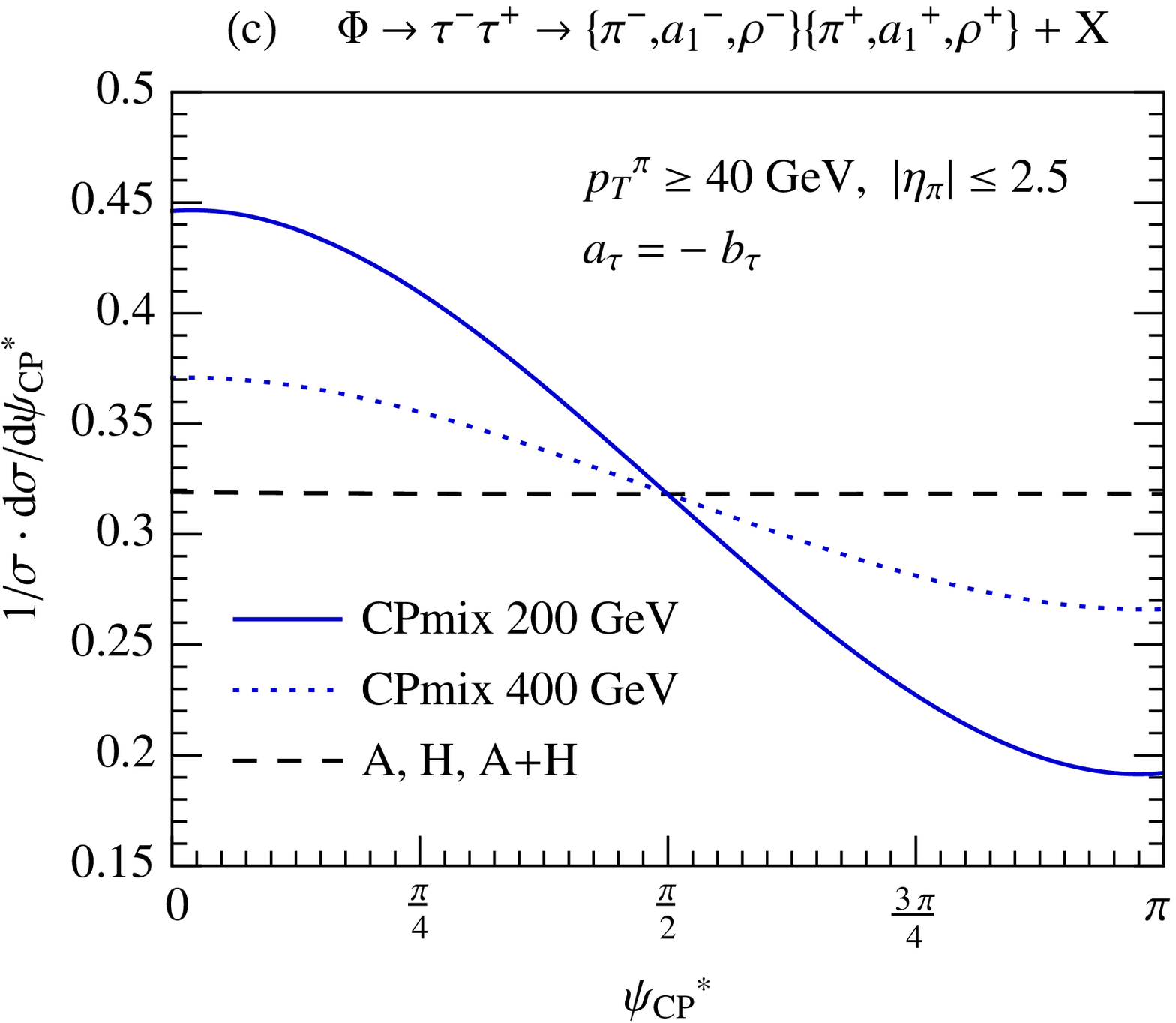}
\caption{
  The $\psi^{*}_{CP}$ distributions for (a) dilepton, (b) 
  lepton-pion, and (c) two-pion final states. The different 
  scenarios are explained in the text.
\label{fig:CPmix_leplep_lephad_phi_ptmi40_eta2.5-1}
}
\end{figure}

The distribution of  $\psi^{*}_{CP}$ is displayed in 
Figs.~\ref{fig:CPmix_leplep_lephad_phi_ptmi40_eta2.5-1}(a) - (c)
for dilepton, lepton-pion, and two-pion final states, respectively.
The figures show that  the variable $\psi^{*}_{CP}$  efficiently
distinguishes between $CP$ conservation and violation -- for the
$CP$-conserving scenarios i) - iii) above, the  distribution  is
flat. For a $CP$-mixed  state, the slope of the   $\psi^{*}_{CP}$ 
distribution for lepton-pion final states is opposite in sign to the
slope for dilepton and for two-pion final states, for reasons
mentioned above. In analogy to (\ref{Aphiasy}) one may consider 
the asymmetry
\smallskip
\begin{equation}
\label{AphiasyCP}                 
A_{\psi^*_{CP}} =  
\frac{N(\psi^*_{CP}>\pi/2) - N(\psi^*_{CP}<\pi/2)}%
{N(\psi^*_{CP}>\pi/2) + N(\psi^*_{CP}<\pi/2)} \, .
\end{equation}
\smallskip
With the values of $A_{\psi^*_{CP}}$ obtained  from the distributions
Figs.~\ref{fig:CPmix_leplep_lephad_phi_ptmi40_eta2.5-1}(a) - (c)
one gets the estimates of the event numbers, given in 
Table~\ref{tab:eventnrCP}, that are needed to find evidence  with 3
s.d.\ significance that   $\Phi$ is an ideal $CP$ mixture. 

\begin{table}[t]
\begin{center}
\begin{tabular}{|c|c|c|c|}
\hline 
~$m_\Phi$ [GeV]~  
    & ~dilepton~
      & ~lepton-pion~
        & ~two-pion~
\tabularnewline
\hline 
200 & 1540 & 514 & 116 
\tabularnewline
400 & 2400 & 1390 & 726
\tabularnewline
\hline
\end{tabular}
\end{center}
\caption{Event numbers needed to find evidence with 3
  s.d.\ significance that $\Phi$ is an ideal $CP$ mixture.
 \label{tab:eventnrCP}
}
\end{table}
 
One may  enhance the discriminating power of the $\psi^{*}_{CP}$ 
distribution by constructing an approximate Higgs-boson rest frame 
as outlined in Sec.~\ref{suse:recoapprox} and impose additional
cuts on the energies of the charged-pions in this frame. In analogy 
to the above results for the $\varphi^{*}$ distribution, we find 
that the improvement as compared to the results shown in
Figs.~\ref{fig:CPmix_leplep_lephad_phi_ptmi40_eta2.5-1}(b), (c) 
is small for Higgs-boson masses below $200$~GeV, whereas it becomes 
significant for heavy $CP$ mixtures with $m_{\Phi} \sim  400$~GeV. 


\section{Conclusions}
\label{sect:concl}

We have shown that the $CP$ quantum numbers of a Higgs-boson 
$\Phi$, produced at the LHC, can be determined with the observables 
(\ref{phistar}) and (\ref{eq:psi_star_distribution}) in the
$\tau$-decay mode $\Phi \to \tau^+ \tau^-$, using all  major 
subsequent 1-prong $\tau$ decays. The selection cuts that we 
applied in our analysis to the dilepton, lepton-pion, and two-pion 
finals states significantly enhance the discriminating power of 
these observables for  the ``non direct charged-pion decay modes''.
Therefore, practically all $\tau$ decay modes can be used
for pinning down the $CP$ properties of  $\Phi$, because 
the three-prong  $\tau$ decays can also be employed for this purpose
\cite{Berge:2008wi}. Depending on the $\Phi$-production cross 
sections, i.e., on its mass
and couplings, it should be feasible to collect the event numbers
(estimated in Tables~\ref{tab:eventnrphi} and~\ref{tab:eventnrCP})
that are required for statistically significant  $CP$ measurements
after several years of high-luminosity runs at the LHC.


\section*{\bf APPENDIX}


In this appendix we collect, for the convenience of the reader,
some results on $\tau$ decays which are relevant for the 
calculations described above. (For a review, see \cite{Stahl:2000aq}.) 
The branching ratios of the 1-prong $\tau$-decay modes,  given in 
Table~\ref{tab:tau_Branching-ratios}, are taken from 
\cite{Nakamura:2010zzi}.

\begin{table}[h]
\begin{tabular}{|c|c|c|c|c|}
\hline 
~decay mode & $\tau^{\pm}\to\pi^{\pm}$~ 
   & ~$\tau^{\pm}\to\rho^{\pm}\to\pi^{\pm}\pi^{0}$~ 
      & ~$\tau^{\pm}\to a_{1}^{\pm}\to\pi^{\pm}2\pi^{0}$~ 
         & ~$\tau^{\pm}\to e^{\pm},\mu^{\pm}$~ 
\tabularnewline
\hline 
$BR{}_{PDG}$ $[\%]$ 
   & $10.91$ 
      & $25.51$ 
         & $9.3$ 
            & $35.2$
\tabularnewline
\hline
\end{tabular}
\caption{Branching ratios for the major 1-prong $\tau$-decay modes
\cite{Nakamura:2010zzi}. 
\label{tab:tau_Branching-ratios}
}
\end{table}

Next we list  the spectral functions $n(E_a)$ and $b(E_a)$ of the
energy-angular distributions (\ref{eq:dGamma_dEdOmega}) of polarized 
$\tau^\mp$ decays to $a^\mp$. The functions given below apply to both 
$\tau^-$ and $\tau^+$ -- but notice the sign change in front of 
$b(E_a)$ in (\ref{eq:dGamma_dEdOmega}). Furthermore, our convention 
for the distribution (\ref{eq:dGamma_dEdOmega}) is such that we
differentiate with respect to the energy $E_a$ of the charged
prong. Therefore, the functions  $n(E_a)$ are dimensionful while
the functions  $b(E_a)$ are dimensionless.


\subsection*{The decay $\tau^\mp \to\pi^\mp +\nu_{\tau}$}

In the 2-body decay $\tau\to\pi+\nu_{\tau}$ the energy $E_{\pi}$ 
in the $\tau$ rest frame is fixed and the functions 
$n_{\pi}(E_{\pi})$ and $b_{\pi}(E_{\pi})$ are given by \cite{Tsai:1971vv}:
\begin{eqnarray}
n_{\pi}(E_{\pi}) 
& = & 
\delta\left(E_{\pi} - \frac{m_{\tau}^2 + m_{\pi}^2}{2m_{\tau}}\right)
\, ,
\qquad
b_{\pi}(E_{\pi})\,\,=\,\,1
\, .
\end{eqnarray}


\subsection*{The decay $\tau^{\mp} \to \rho^{\mp} \to 
\pi^{\mp}\pi^{0} \nu_{\tau}$}

The differential rate of the decay of polarized  $\tau$ leptons to a
charged  pion via a $\rho$-meson was calculated in \cite{Tsai:1971vv}.
With $x = 4E_{\pi} / m_{\tau}$, where $E_{\pi}$ denotes the energy of
the charged-pion in  the $\tau$ rest frame, the spectral functions are 
given by
\begin{eqnarray*}
n_{\rho}(E_{\pi}) 
& = &  \frac{6}{m_{\tau}}
\frac{(x-r-1)^{2}+(1-r)(r-p)}{(1-r)^{2}(1+2r)(1-p/r)^{3/2}}
\, ,
\\
b_{\rho}(E_{\pi}) 
& = & 
\frac{x(x-r-1)^{2}+x(3-r)(r-p)-4(r-p)}%
{\sqrt{x^{2}-4p}\,((x-r-1)^{2}+(1-r)(r-p))}
\, , 
\end{eqnarray*}
with $p = 4m_{\pi}^{2} / m_{\tau}^{2}$ and $r = m_{\rho}^{2} / 
m_{\tau}^{2}$. These functions are plotted in 
Fig.~\ref{fig:a1_rho_nx_bx}(a). The kinematic range of  $E_{\pi}$ is 
\begin{equation}
\frac{m_{\tau}}{4}
\left(1+r-(1-r)\sqrt{1-\frac{p}{r}}\right)
\le 
E_{\pi}
\le
\frac{m_{\tau}}{4}
\left(1+r+(1-r)\sqrt{1-\frac{p}{r}}\right) \, .
\end{equation} 


\subsection*{The decay $\tau^{\mp} \to a_{1}^{\mp} \to 
\pi^{\mp}2\pi^{0}\ \nu_{\tau}$}

The differential rate of the 1-prong decay of polarized $\tau$ 
leptons to a charged-pion via a $a_{1}$-meson was  calculated in 
\cite{Overmann:1992zv}. The corresponding functions $n_{a_1}(E_{\pi})$ 
and $b_{a_1}(E_{\pi})$ are complicated and were fitted to the 
numerical results shown in Fig.~6-4 of Ref.\ \cite{Overmann:1992zv}. 
With 
\begin{equation}
x  = 
\frac{2m_{\tau}(E_{\pi}-m_{\pi})}{m_{\tau}^{2}-3m_{\pi}^{2} 
- 2m_{\tau} m_{\pi}}
\end{equation}
where $E_{\pi}$ is the energy of the charged-pion in the $\tau$ rest 
frame, we obtain 
\begin{eqnarray}
n_{a_1}(E_{\pi}) 
& = & 
\frac{2m_{\tau}}{m_{\tau}^2 - 3m_{\pi}^2 - 2m_{\tau} m_{\pi}}
\Bigl(0.0112624 - 2.15495 x + 165.368 x^{2}
\nonumber \\
&  & 
- 997.586 x^{3} + 2818.75 x^{4} - 4527.77 x^{5}
\\
&  & 
+ 4250.43 x^{6} - 2182.33 x^{7} + 475.283 x^{8}
\Bigr) \, , 
\nonumber \\[1ex]
b_{a_1}(E_{\pi}) 
& = & 
-5.28726 \sqrt{x} + 9.38612 x - 1.26356 x^{2}
\\
&  & 
-18.9094 x^{3} + 36.0517 x^{4} - 19.4113 x^{5} \, .
\nonumber
\end{eqnarray}
The plots of  $n_{a_1}(E_{\pi})$ and $b_{a_1}(E_{\pi})$ are shown in 
Fig.~\ref{fig:a1_rho_nx_bx}(b). The kinematical range of the 
charged-pion energy in the $\tau$ rest frame is
\begin{equation}
m_{\pi}
\le 
E_{\pi}
\le
\frac{m_{\tau}^{2} - 3m_{\pi}^{2}}{2m_{\tau}}\, .
\end{equation}


\subsection*{The decay $\tau^{\mp} \to l^{\mp} \nu_{l} \nu_{\tau}$}

For the leptonic decays $\tau^{\pm} \to l^{\pm} \nu_{l} \nu_{\tau}$ 
the mass of the final state lepton, $e$ or $\mu$, can be 
neglected. Using $x = 2E_{l} / m_{\tau}$, where  $E_{l}$ is 
defined in the $\tau$ rest frame, one has \cite{Tsai:1971vv}
\begin{eqnarray}
n_l(E_{l}) 
& = & 
\frac{4}{m_{\tau}} x^{2}\,\left(3-2x\right)
\, ,
\qquad
b_l(E_{l})
=
\frac{1-2\, x}{3-2\, x}
\label{eq:lep_nx_bx}
\end{eqnarray}
with $0 \le E_{l} \le m_{\tau} / 2$. 


\section*{Acknowledgments}

The work of S.~B.\ is supported by the Initiative and Networking 
Fund of the Helmholtz Association, Contract No. HA-101 (``Physics at 
the Terascale'') and by the Research Center ``Elementary Forces and 
Mathematical Foundations'' of the Johannes-Gutenberg-Universit\"at 
Mainz. The work of W.~B.\ is supported by BMBF.



\begin{thebibliography}{99} 
 
\bibitem{Berge:2008dr}
  S.~Berge and W.~Bernreuther,
  Phys.\ Lett.\  {\bf B671 }, 470 (2009).
  [arXiv:0812.1910 [hep-ph]].

\bibitem{Djouadi:2005gi} 
  A.~Djouadi, 
  Phys. Rept. {\bf 457}, 1 (2008);
  [arXiv:hep-ph/0503172].

\bibitem{Djouadi:2005gj}
  A.~Djouadi, Phys. Rept. {\bf 459}, 1 (2008); [arXiv:hep-ph/0503173].



\bibitem{GomezBock:2007hp}
  M.~Gomez-Bock, M.~Mondragon, M.~M\"uhlleitner, M.~Spira and 
  P.~M.~Zerwas,
  ``Concepts of Electroweak Symmetry Breaking and Higgs Physics,''
  [arXiv:0712.2419 [hep-ph]].

\bibitem{Grojean:2007zz}
  C.~Grojean,
  Phys.\ Usp.\  {\bf 50}, 1 (2007).

\bibitem{Morrissey:2009tf}
  D.~E.~Morrissey, T.~Plehn and T.~M.~P.~Tait,
  ``Physics searches at the LHC,''
  [arXiv:0912.3259 [hep-ph]].

\bibitem{Dell'Aquila:1985ve} 
  J.~R.~Dell'Aquila and C.~A.~Nelson, 
  Phys.\ Rev.\  D {\bf 33}, 80 (1986).
 
\bibitem{Dell'Aquila:1988fe} 
  J.~R.~Dell'Aquila and C.~A.~Nelson, 
  Nucl.\ Phys.\  B {\bf 320}, 86 (1989). 

\bibitem{Bernreuther:1993df} 
  W.~Bernreuther and A.~Brandenburg, 
  Phys.\ Lett.\  B {\bf 314}, 104 (1993);
  Phys.\ Rev.\  D {\bf 49}, 4481 (1994). 

\bibitem{Chang:1993jy}
  D.~Chang, W.~Y.~Keung and I.~Phillips, 
  Phys.\ Rev.\  D {\bf 48}, 3225 (1993).

\bibitem{Soni:1993jc}
  A.~Soni, R.~M.~Xu,
  Phys.\ Rev.\  {\bf D48}, 5259 (1993).
  [hep-ph/9301225].
 
\bibitem{Kramer:1993jn}
   M.~Kr\"amer, J.~H.~K\"uhn, M.~L.~Stong and P.~M.~Zerwas, 
  Z.\ Phys.\  C {\bf 64}, 21 (1994).

\bibitem{Grzadkowski:1995rx}
  B.~Grzadkowski and J.~F.~Gunion, 
  Phys.\ Lett.\  B {\bf 350}, 218 (1995). 

\bibitem{Bernreuther:1997af}
  W.~Bernreuther, A.~Brandenburg and M.~Flesch, 
  Phys.\ Rev.\  D {\bf 56}, 90 (1997). [hep-ph/9701347].

\bibitem{Plehn:2001nj}
  T.~Plehn, D.~L.~Rainwater and D.~Zeppenfeld, 
  Phys.\ Rev.\ Lett.\  {\bf 88}, 051801 (2002). 

\bibitem{Buszello:2002uu}
  C.~P.~Buszello {\it et al.}, 
  Eur.\ Phys.\ J.\  C {\bf 32}, 209 (2004). 

\bibitem{Klamke:2007cu}
  G.~Klamke and D.~Zeppenfeld,
  JHEP {\bf 0704}, 052 (2007).
  [hep-ph/0703202 [HEP-PH]].

\bibitem{Berge:2008wi}
  S.~Berge, W.~Bernreuther and J.~Ziethe,
  Phys.\ Rev.\ Lett.\  {\bf 100}, 171605 (2008).
  [arXiv:0801.2297 [hep-ph]].

\bibitem{DeRujula:2010ys}
  A.~De Rujula, J.~Lykken, M.~Pierini, C.~Rogan and M.~Spiropulu,
  Phys.\ Rev.\  {\bf D82}, 013003 (2010).
  [arXiv:1001.5300 [hep-ph]].


\bibitem{Accomando:2006ga}
  E.~Accomando {\it et al.},
  ``Workshop on CP Studies and Nonstandard Higgs Physics,''
  [hep-ph/0608079].
 

\bibitem{Baglio:2010ae}
  J.~Baglio, A.~Djouadi,
  JHEP {\bf 1103}, 055 (2011).
  [arXiv:1012.0530 [hep-ph]].


\bibitem{Baglio:2011xz}
 J.~Baglio, A.~Djouadi,
 [arXiv:1103.6247 [hep-ph]].


\bibitem{Chatrchyan:2011nx}
  S.~Chatrchyan {\it et al.} [CMS Collaboration],
  Phys.\  Rev.\  Lett.\  {\bf 106}, 231801 (2011).
  [arXiv:1104.1619 [hep-ex]];
 CMS Collaboration, CMS PAS HIG-11-009.
  
\bibitem{Collaboration:2011rv}
G.~Aad {\it et al.}  [ATLAS Collaboration],
    Phys.\ Lett.\ B {\bf 705}, 174 (2011).

 
\bibitem{Harlander:2007zz}
  R.~Harlander,
  J.\ Phys.\ G {\bf G35}, 033001 (2008).
  
\bibitem{Dittmaier:2011ti}
  S.~Dittmaier {\it et al.} [LHC Higgs Cross Section Working Group 
  Collaboration],
  ``Handbook of LHC Higgs Cross Sections: 1. Inclusive Observables,''
  [arXiv:1101.0593 [hep-ph]].

\bibitem{Ellis:1987xu}
  R.~K.~Ellis, I.~Hinchliffe, M.~Soldate and J.~J.~van der Bij,
  Nucl.\ Phys.\  {\bf B297}, 221 (1988).

\bibitem{Elagin:2010aw}
  A.~Elagin, P.~Murat, A.~Pranko and A.~Safonov,
Nucl. Instrum. Methods Phys. Res., Sect. A 654, 481
(2011).
  [arXiv:1012.4686 [hep-ex]]. 


\bibitem{Perchalla2011}
L. Perchalla,  Ph.D. thesis,
``Kinematic Tau Reconstruction and Search For The Higgs Boson in
Hadronic Tau Pair Decays with the CMS Experiment'',
 RWTH Aachen (2011).


\bibitem{Bernreuther:1989kc}
  W.~Bernreuther, O.~Nachtmann,
  Phys.\ Rev.\ Lett.\  {\bf 63}, 2787 (1989).

\bibitem{Bernreuther:1993nd}
  W.~Bernreuther, O.~Nachtmann, P.~Overmann,
  Phys.\ Rev.\  {\bf D48}, 78 (1993).

\bibitem{Pierzchala:2001gc}
  T.~Pierzchala, E.~Richter-Was, Z.~Was, M.~Worek,
  Acta Phys.\ Polon.\  {\bf B32}, 1277 (2001).
  [hep-ph/0101311].


\bibitem{Rosendahl:2011si}
  P.~L.~Rosendahl, T.~Burgess, B.~Stugu,
  [arXiv:1105.6003 [hep-ex]].




\bibitem{Cohen:1993nk} 
  A.~G.~Cohen, D.~B.~Kaplan and A.~E.~Nelson, 
  Ann.\ Rev.\ Nucl.\ Part.\ Sci.\  {\bf 43}, 27 (1993). 
 
\bibitem{Bernreuther:2002uj}
  W.~Bernreuther,
  Lect.\ Notes Phys.\  {\bf 591}, 237 (2002).
  [hep-ph/0205279].

\bibitem{Stahl:2000aq} 
  A.~Stahl, 
  Springer Tracts Mod.\ Phys.\  {\bf 160}, 1 (2000). 
 
\bibitem{Nakamura:2010zzi}
  K.~Nakamura {\it et al.} [Particle Data Group Collaboration],
  J.\ Phys.\ G {\bf G37}, 075021 (2010).

\bibitem{Tsai:1971vv}
  Y.~-S.~Tsai,
  Phys.\ Rev.\  {\bf D4}, 2821 (1971); Erratum-ibid. {\bf D13}, 
  771 (1976).

\bibitem{Overmann:1992zv}
  P. Overmann, Ph.D. thesis,
  Universit\"at
  Heidelberg, 1992, preprint HD-THEP-92-38.





\end{thebibliography}
\end{document}